\pdfoutput=1
\documentclass[%
 aip,
 pop,
 amsmath,amssymb,
 reprint,%
]{revtex4-1}

\usepackage{graphicx}
\usepackage{bm}
\usepackage{dcolumn}

\usepackage{hyperref}
\usepackage{amsmath}

\usepackage[utf8]{inputenc}
\usepackage[T1]{fontenc}
\usepackage{mathptmx}

\newcommand*\dif{\mathop{}\!\mathrm{d}}
\newcommand*\mathth{^{\text{th}}}
\usepackage{fix-cm} 
\usepackage{siunitx} 
\usepackage{xfrac} 
\usepackage[disable]{todonotes}
\usepackage{ifthen} 

\usetikzlibrary{matrix,arrows.meta, fit, positioning,calc}
\tikzstyle{neuron}=[circle,fill=black!25,minimum size=17pt,inner sep=0.5pt]
\tikzstyle{input neuron}=[neuron, fill=green!50]
\tikzstyle{output neuron}=[neuron, fill=red!50]
\tikzstyle{hidden neuron}=[neuron, fill=blue!50]
\tikzstyle{annot} = [text width=4em, text centered]
\def\layersep{2.5}
\def\nodedist{0.8}
\def\blocksep{1}
\tikzset{
  pics/ffnn/.style args={#1 inputnodes #2 hiddenlayers #3 hiddennodeblocks #4 outputnodes #5 outputlabel #6 twoblock #7}{
    code={
      \def\prefix{#1}
      \def\inputnodes{#2}
      \def\hiddenlayers{#3}
      \def\hiddennodeblocks{#4}
      \def\outputnodes{#5}
      \def\outputlabel{#6}
      \def\twoblock{#7}
      \pgfmathparse{(1+0.75*(\hiddenlayers-1))*\layersep}\let\lastlayerxcoord=\pgfmathresult
      \pgfmathparse{-(\hiddennodeblocks * \nodedist)}\let\lastnodeblockycoord=\pgfmathresult
      \pgfmathparse{\hiddennodeblocks+1}\let\secondblockstart=\pgfmathresult
      \ifthenelse{\equal{\twoblock}{1}} {
        \pgfmathparse{2*\hiddennodeblocks}\let\secondblockend=\pgfmathresult
      }{
        \let\secondblockend=\hiddennodeblocks
      }
      \foreach \name / \y in {1,...,\inputnodes}
      \node[input neuron,  pin={[pin edge={<-}]180:$x_\y$}] (\prefix-I-\name) at (0,-\y) {};

      \foreach \layernum in {1,...,\hiddenlayers} {
        \pgfmathparse{(1+0.75*(\layernum-1))*\layersep}\let\layerxcoord=\pgfmathresult
        \foreach \yy in {1,...,\hiddennodeblocks}
        \pgfmathparse{-(\yy * \nodedist)}\let\nodeycoord=\pgfmathresult
        \node[hidden neuron] (\prefix-H\layernum-\yy) at (\layerxcoord,\nodeycoord) {};

        \ifthenelse{\equal{\twoblock}{1}} {
          \foreach \yy in {0,...,2}
          \pgfmathparse{\lastnodeblockycoord - 0.3 * \yy - 0.6}\let\dotycoord=\pgfmathresult
          \draw[fill=black] (\layerxcoord, \dotycoord) circle[radius=0.08];

          \foreach \yy [evaluate=\yy using int(\yy)] in {\secondblockstart,...,\secondblockend}
          \pgfmathparse{-(\yy * \nodedist + \blocksep)}\let\nodeycoord=\pgfmathresult
          \node[hidden neuron] (\prefix-H\layernum-\yy) at (\layerxcoord,\nodeycoord) {};
        }
      }

       \ifthenelse{\equal{\outputlabel}{""}} {
         \node[output neuron] (\prefix-O) at (\lastlayerxcoord+\layersep, \lastnodeblockycoord-0.9) {};
       }{
         \node[output neuron,pin={[pin edge={->}]right:\outputlabel}] (\prefix-O) at (\lastlayerxcoord+\layersep, \lastnodeblockycoord-0.9) {};
       }

      \foreach \source [evaluate=\source using int(\source)] in {1,...,\inputnodes}
      \foreach \dest[evaluate=\dest using int(\dest)] in {1,...,\secondblockend}
      \path (\prefix-I-\source.east) edge (\prefix-H1-\dest.west);

      \foreach \source [evaluate=\source using int(\source)] in {1,...,\secondblockend}
      \foreach \dest[evaluate=\dest using int(\dest)] in {1,...,\secondblockend}
      \foreach \layernum [remember=\layernum as \prevlayernum (initially 1)] in {2,...,\hiddenlayers}
      \path (\prefix-H\prevlayernum-\source.east) edge (\prefix-H\layernum-\dest.west);

      \foreach \source [evaluate=\source using int(\source)] in {1,...,\secondblockend}
      \path (\prefix-H\hiddenlayers-\source.east) edge (\prefix-O.west);
    }
  }
}
\graphicspath{{Pictures/}}

\begin{document}


\title{Fast modelling of turbulent transport in fusion plasmas using neural networks} 



\newcommand{\affdiffer}{DIFFER, PO Box 6336, 5600 HH Eindhoven, The Netherlands}
\author{K.L. van de Plassche}
\email[]{k.l.vandeplassche@differ.nl}
\affiliation{\affdiffer}
\author{J. Citrin}
\affiliation{\affdiffer}
\author{C. Bourdelle}
\affiliation{CEA, IRFM, F-13108 Saint-Paul-lez-Durance, France}
\author{Y. Camenen}
\affiliation{CNRS, Aix-Marseille Univ., PIIM UMR7345, Marseille, France}
\author{F. J. Casson}
\affiliation{CCFE, Culham Science Centre, OX14 3DB, Abingdon, UK}
\author{V.I. Dagnelie}
\affiliation{\affdiffer}
\affiliation{ITP, Utrecht University, Princetonplein 5, 3584 CC Utrecht, The Netherlands}
\author{F. Felici}
\affiliation{EPFL-SPC, CH-1015, Lausanne, Switzerland}
\author{A. Ho}
\affiliation{\affdiffer}
\author{S. Van Mulders}
\affiliation{EPFL-SPC, CH-1015, Lausanne, Switzerland}
\author{JET Contributors}
\affiliation{See the author list of  E. Joffrin et al. accepted for publication in Nuclear Fusion Special issue 2019, https://doi.org/10.1088/1741-4326/ab2276}
\noaffiliation


\date{\today}

\begin{abstract}
We present an ultrafast neural network (NN) model, QLKNN, which predicts core tokamak transport heat and particle fluxes. QLKNN is a surrogate model based on a database of 300 million flux calculations of the quasilinear gyrokinetic transport model QuaLiKiz. The database covers a wide range of realistic tokamak core parameters. Physical features such as the existence of a critical gradient for the onset of turbulent transport were integrated into the neural network training methodology. We have coupled QLKNN to the tokamak modelling framework JINTRAC and rapid control-oriented tokamak transport solver RAPTOR. The coupled frameworks are demonstrated and validated through application to three JET shots covering a representative spread of H-mode operating space, predicting turbulent transport of energy and particles in the plasma core. JINTRAC-QLKNN and RAPTOR-QLKNN are able to accurately reproduce JINTRAC-QuaLiKiz $T_{i,e}$ and $n_e$ profiles, but 3 to 5 orders of magnitude faster. Simulations which take hours are reduced down to only a few tens of seconds. The discrepancy in the final source-driven predicted profiles between QLKNN and QuaLiKiz is on the order 1\%-15\%. Also the dynamic behaviour was well captured by QLKNN, with differences of only 4\%-10\% compared to JINTRAC-QuaLiKiz observed at mid-radius, for a study of density buildup following the L-H transition. Deployment of neural network surrogate models in multi-physics integrated tokamak modelling is a promising route towards enabling accurate and fast tokamak scenario optimization, Uncertainty Quantification, and control applications.
\end{abstract}


\maketitle 

\section{Introduction}
Accurate prediction of tokamak core plasma temperature and density is essential for interpretation and preparation of current-day fusion experiments, optimization of plasma scenarios, and designing future devices. Time-evolved tokamak simulation on discharge timescales is typically carried out within an 'integrated modelling' approach\cite{poli2018_integrated_modeling}, where multiple models representing various physics phenomena are coupled together within a single code or workflow. An essential component of integrated models is the prediction of turbulent fluxes, particularly in the tokamak core where transport is often dominated by plasma microinstabilities\cite{doyle2007_turbulence_iter,horton2003_turbulence_basics}. However, calculating these fluxes using nonlinear gyrokinetic models is too computationally expensive for routine simulation of tokamak discharge evolution.

Reduced order turbulence models have thus been developed for increased tractability. They remain first-principle-based yet are computationally cheaper through invoking the quasilinear approximation. Quasilinear turbulence models like QuaLiKiz\cite{citrin2017_qualikiz,bourdelle2016_qualikiz_benchmark,qualikiz_website} and TGLF\cite{staebler2007_tglf} are valid in wide parameter regimes in the tokamak core and have been extensively validated against nonlinear gyrokinetics\cite{casati2009_quasilinear,citrin2012_qualikiz_mshear,cottier2013_quasilinear}. These models can predict turbulent fluxes approximately 6 orders of magnitude faster than $\delta f$ local nonlinear codes. For QuaLiKiz, this means around 10 minutes for a radial profile of 25 multiscale transport fluxes on a single CPU, depending on the physics fidelity used in the simulation. The speed has enabled routine runs of QuaLiKiz coupled to integrated modelling suites such as JINTRAC\cite{cenacchi1988_jetto,romanelli2014_jintrac}, recently leading to numerous successful validation exercises against JET\cite{citrin2017_qualikiz,ho2019_gp,breton2018_qualikiz,casson2018_92398} and AUG\cite{linder2019_flux} discharges. However, due to the small timestep needed in explicit transport PDE solvers, these integrated models need thousands of calls to the turbulent transport module for each second of plasma evolution, and thus can still can take days to run, even when parallelized. This sets limits on large-scale model validation and theory-based optimization of fusion experiments, as well as for model-based real-time control applications.

To further accelerate integrated modelling workflows we apply feed-forward neural networks (FFNNs) as a surrogate model, reproducing the underlying turbulent transport model within tens of microseconds. The concept takes advantage of the fast evaluation time of the reduced tokamak turbulence models (e.g. QuaLiKiz), by applying them for generating large training sets then used for neural network regression. The neural networks can then be used as a drop-in replacement inside the integrated model, removing one of the main computational bottlenecks.

Similar development of neural network surrogates for physics models applied within tokamak integrated modelling has been carried out for: the TGLF quasilinear turbulent transport model\cite{meneghini2017_core-pedestal-neural-network}, the EPED pedestal confinement model\cite{meneghini2017_core-pedestal-neural-network} and the neutral beam heating code NUBEAM\cite{goldston1981_nubeam,pankin2004_nubeam,boyer2019_nubeamnet}. This paper presents the state-of-the-art of the QuaLiKiz neural network surrogate model, far beyond our original proof-of-principle\cite{citrin_nn_proof_concept,felici2018_raptorqlknn4d}. In the original proof-of-principle, neural networks were trained on a small dataset and implemented into the control-oriented tokamak simulator RAPTOR\cite{felici2018_raptorqlknn4d}. It was shown that this method could accurately predict the temperature profiles of a JET discharge, giving confidence to apply this methodology on a larger scale. In this work, we extend the input dimensionality of the model from 4 to 10, leading to significantly increased surrogate model fidelity. We chose to generate the QuaLiKiz training set as a large regular input parameter hyperrectangle scan (see section \ref{sec:dataset}), to ensure wide applicability of the obtained model. Since neural networks extrapolate poorly beyond training dataset bounds, a large and experimentally relevant database is essential for good model performance.

Other novel aspects include a focus on incorporating physics-based features in the training pipeline, as discussed in sections \ref{sec:physics_based_nn} and \ref{sec:pipeline}. To properly introduce the applied methodology, we summarize neural network techniques in section \ref{sec:neural_networks}. We show a new analytical scaling rule partially capturing the effect of rotation on transport fluxes in section \ref{sec:rotation_rule}. Finally, we show the application of the neural network surrogate model within the control-oriented transport code RAPTOR and the integrated modelling framwork JINTRAC in section \ref{sec:transport_codes}.
%
%
%
\section{Dataset generation}
\label{sec:dataset}
We use the quasilinear gyrokinetic transport model QuaLiKiz to generate a large database of turbulent transport model calculations. QuaLiKiz solves the linear gyrokinetic dispersion relation in the electrostatic limit in $s-\alpha$ geometry. By assuming a shifted Gaussian for the mode eigenfunctions in the strong ballooning limit, strongly trapped and passing particles, and a small Mach number, the calculation is greatly simplified leading to increased calculation speed ($\times10^3$) beyond standard linear gyrokinetics. The quasilinear approximation is then used, setting the transport fluxes (heat, particle, and momentum) from the linear response over a range of wavevectors, in conjunction with a \textit{saturation rule} for the electrostatic potential amplitudes and spectral shape, tuned to nonlinear gyrokinetic simulations both at ion-scales and electron-scales\cite{bourdelle2016_qualikiz_benchmark,citrin2017_gene}.

The input space of the full QuaLiKiz code ($\sim$15 dimensions for typical simulations) is too large to cover with a brute-force hypercube scan. We thus constrain the training set dimensionality to the subset most significantly impacting turbulent transport within the framework of QuaLiKiz approximations. These input dimensions are shown in Table \ref{tab:9d_input_space} and include the logarithmic ion temperature gradient ($R/L_{T_i}$), electron temperature gradient ($R/L_{T_e}$), density gradient ($R/L_n$), ion-electron temperature ratio ($T_i/T_e$), safety factor ($q$), magnetic shear ($\hat{s}$), local inverse aspect ratio ($r/R$), collisionality ($\nu^*$), and effective charge ($Z_{eff}$), with a carbon impurity and deuterium main ion. The impurity ion density is controlled by $Z_{eff}$, scanning it independently from $\nu^{*}$. Notable simplifications are excluding plasma rotation ($\gamma_{E\times B} = v_{par} = v_{perp} = 0$), assuming equal density gradient for the two ion species, and no Shafranov shift. This significantly extends the previous proof-of-principle 4D neural network QuaLiKiz regression \cite{citrin_nn_proof_concept}, which included only $q$, $\hat{s}$, $R/L_{T_i}$, and $T_i/T_e$ as input. The nine inputs are taken as the feature space of the neural network. The impact of rotation, important for accurate tokamak plasma simulation, is taken into account through a new separate model in post-processing, as described in Section \ref{sec:rotation_rule}.

A  database consisting of $3 \times 10^8$ QuaLiKiz input-flux relations was generated with HPC resources on the Edison supercomputer at NERSC, using \SI{1.3}{MCPUh}. The database spans ion scales ($k_\theta \rho_s \leq 2$) and electron scales ($k_\theta \rho_s > 2$) and contains contributions to transport fluxes and coefficients $q$ (heat), $\Gamma$ (total particle), $D$ (particle diffusivity), and $V$ (particle convection) per species. The input space was chosen as a rectangular, non-uniform 9-dimensional grid. The bounds cover dimensionless parameter regimes typically encountered in the core of standard aspect-ratio present-day tokamaks, and future devices such as ITER and DEMO. We chose the spacing of the grid to have a higher density around typical threshold zones (e.g. $-\frac{R}{T_{e}}\frac{\dif T_{e}}{\dif r} \approx 5$) and  zones of high non-monoticity (e.g. $\hat{s} \approx 0.7$) based on previous extensive experience with application of QuaLiKiz within integrated modelling and standalone. See Table \ref{tab:9d_input_space} for the bounds of the generated dataset. The dataset is stored in HDF5(pandas) or netCDF(xarray) format, takes around 12 GiB for the 9D input set, and 3 GiB per output variable, and is freely available on Zenodo\cite{zenodo_dataset10D}.
\begin{table}
  \caption{\label{tab:9d_input_space}9D hyperrectangle bounds and number of points of the QuaLiKiz neural network training set. Each input is non-uniformly distributed in space, with a finer resolution in experimentally more relevant regimes.}
  \begin{ruledtabular}
    \begin{tabular}{*{1}{l} *{3}{r}}
      variable & \# points & min & max \\
      \hline
      $k_\theta \rho_s \leq 2$ & 10 & 0.1 & 2 \\
      $k_\theta \rho_s > 2$ & 8 & 3.5 & 36 \\
      \hline
      $R / L_{T_e}$ & 12 & 0 & 14 \\
      $R / L_{T_i}$ & 12 & 0 & 14 \\
      $R / L_n$ & 12 & -5 & 6 \\
      $q$ & 10 & 0.66 & 15\\
      $\hat{s}$ & 10 & -1 & 5\\
      $r / R$ & 8 & 0.03 & 0.33 \\
      $T_i / T_e$ & 7 & 0.25 & 2.5 \\
      $\nu^*$ & 6 & $1 \times 10^{-5}$ & 1 \\
      $Z_{eff}$ & 5 & 1 & 3 \\
      \hline\hline
      Total flux calculations & $3 \times 10^8$ & $\approx \SI{1.3}{MCPUh}$ &
    \end{tabular}
  \end{ruledtabular}
\end{table}

\section{Rotation rule}
\label{sec:rotation_rule}
To save computation time, the dataset was ran without rotation. Beyond adding additional dimensions in training set input-space with associated cost, running QuaLiKiz with rotation takes $\times4$ more computation time due to a loss of symmetry in the internal integration routines. However, since the impact of rotation on confinement can be critical, particularly in high performance H-modes, we implemented a new flux suppression rule in postprocessing. This rule is based on a new set of linear GENE\cite{jenko2000_gene} scans around the GA-Standard case, coupled to a methodology to assess the impact of rotation on linear growth-rates in spite of the Floquet fluctuations \cite{dagnelie2017_thesis,dagnelie_itg_flow_shear}. These scans consisted of toroidal rotation scans for various $q$, $\epsilon\equiv\frac{r}{R}$, and $\hat{s}$, capturing both the effects of $E\times B$ stabilisation and Parallel Velocity Gradient (PVG) destabilization. The rule scales all ion-scale fluxes with a tuned function $f_{rot}(q, \hat{s}, \epsilon)$. It depends also on the rotationless maximum ion-scale growth rate $\gamma_0$, which is predicted by an additional neural network based on the HPC-generated QuaLiKiz database, and the normalized $E\times B$ shearing rate $\gamma_{E\times B}$ defined in Equation \ref{eq:gammaE}.
\begin{align}
  \label{eq:gammaE}
  \gamma_{E\times B} &\equiv - \frac{\dif v_{perp}}{\dif r} \frac{R}{c_{ref}} \\
  c_{ref} &\equiv \sqrt{\frac{T_{ref}}{m_p}}
\end{align}
Where $v_{perp}$ is the $E\times B$ velocity, $T_{ref}$ is a reference temperature of \SI{1}{keV}, and $m_p$ is the proton mass. The TEM/ITG ion $i$ and electron $e$ transport coefficient $x$ is then scaled with $f_{rot}$ as described in Equation \ref{eq:rotrule}.
\begin{align}
  \label{eq:rotrule}
    f_{rotrule} &= c_1 q + c_2 \hat{s} + c_3 / \epsilon - c_4 \\
    f_{rot} &= \max(1 + f_{rotrule} \gamma_E / \gamma_0, 0) \\
    x_{i/e, ITG/TEM} &= f_{rot} * x_{i/e, ITG/TEM} 
\end{align}
Where the values of the constants were determined to be $c_1 = 0.13$, $c_2 = 0.41$, $c_3 = 0.09$, and $c_4 = 1.65$. Using this rule, we are able to capture partially the effect of rotation on transport in a computationally quick way.
\section{Physics-based neural network training}
\label{sec:physics_based_nn}
Regularized neural networks, described in detail in Appendix \ref{sec:neural_networks}, provide a smooth regression of supplied training data. It does not assume any features of the underlying mapping. Physics-informed features can be directly implemented into the training methodology to significantly improve the fidelity of the surrogate transport model. For our application, we desire the following features:
\begin{itemize}
    \item sharp flux discontinuities at critical (temperature) gradients of the underlying instabilities
    \item identical critical (temperature) gradient for all transport channels driven by a single (TEM/ITG) instability
\end{itemize}
This was found essential for consistent results in integrated modelling. We show an example of a physics-unaware model in integrated modelling in Section \ref{sec:naive}.

We can include these physical features in the neural network training process itself. This is done by choosing training targets carefully, resulting in \textit{leading flux} predicting networks, and \textit{flux ratio} predicting networks for each turbulence driving mode in QuaLiKiz, see Appendix \ref{sec:mode_split}. For example, for ITG we predict a leading ion flux $q_{i,ITG}$ and flux ratios $q_{e,ITG}/q_{i,ITG}$ and $\Gamma_{e,ITG}/q_{i,ITG}$. When combined, the threshold is necessarily at the same location in 9D input space. This method is described in more detail in Appendix \ref{sec:training_targets}. The predicted fluxes are in normalized GyroBohm units, summarized in Appendix \ref{sec:gyrobohm}. Since the underlying physics model follows GyroBohm scaling, developing the neural network regression for dimensionless quantities allows a reduction in input dimensions. Conversion of outputs to SI quantities is carried out in post-processing as per the GyroBohm scaling.

The sharpness of threshold can be improved by modifying the cost function optimized during network training. We describe two such modifications in Appendix \ref{sec:cost_function}. Firstly, we calculate the classical goodness-of-fit term, in this work the mean square error, only on the turbulent unstable points. Secondly, we add a term that punishes neural network predictions in the turbulent stable zone. These modifications could be easily included in our training pipeline, which is described in Appendix \ref{sec:pipeline}.

As with all data-driven approaches, the quality of the data used is paramount for resulting model performance. In this work, a conservative approach to data filtering was used. All untrusted and unvalidated regime QuaLiKiz data is filtered out. Most notably, as described in Appendix \ref{sec:filtering}, we disregard regions with negative heat fluxes. The TEM fluxes might also be filtered too strictly, resulting in worse performance for these networks. The filtering can be improved with more rigorous analysis of the dataset, and better validation of QuaLiKiz in unexplored regimes, both left for future work.

It was found that to judge performance of the resulting surrogate model inside transport codes, classical machine learning measures of goodness were not sufficient. Classically, one looks at global performance, e.g. total root mean square error between network and original data, or global smoothness. In our work, we instead found local and physical features to be more important. While no definite metric for goodness was found, we show measurable performance indicators found important in this work in Appendix \ref{sec:measures}.
We applied this methodology to train thousands of neural networks in a hyper-parameter tuning exercise. All networks were judged against the aforementioned performance indicators measured against a separate test data set, to show generalization capability on unseen data. The optimization resulted in a final set of neural networks with the hyper-parameters shown in Table \ref{tab:hyperpar}. These scored well on global root mean square error and other parameters, as shown in Table \ref{tab:measures}. The networks are implemented within integrated modelling suites for transport physics validation, using an open-source Fortran driver, as will be shown in section \ref{sec:transport_codes}. The speed of the integrated modelling is vastly accelerated, which is the primary motivation of this study. The neural networks and fast driver evaluate a 24-point radial profile of transport coefficients within 1.4 ms without derivatives, and 60 ms with derivatives, on a single core. 
\section{Application in transport codes}
\label{sec:transport_codes}
 The time evolution of the radial profiles of plasma current, density, temperatures, and angular momentum, is governed by a highly non-linear, coupled system of partial differential equations, describing radial transport in the plasma core. Generally this system is too complicated to solve fully explicitly from first principles with direct numerical simulation, so assumptions have to be made to improve tractability. Timescale separation between transport and turbulent process timescales allows the system of equations to be decoupled in mathematically and computationally decoupled \textit{modules}. This is illustrated in the 1D energy equation (in cylindrical coordinates for simplicity) shown in Equation \ref{eq:transport}. Analogues exist for the poloidal magnetic flux diffusion equation, density equation and momentum equation. The transport flux $q_s$ and sources $Q_s(r, t)$ are typically calculated by physics models, under the assumption that the process timescale is much less than the representative transport timescale, e.g. the particle and energy confinement time for temperature and density profile evolution. In this study we focus on the energy and density transport, as these are the coefficients calculated by QLKNN.
\begin{align}
  \label{eq:transport}
  \frac{3}{2} \frac{\left(\partial n_s T_s \right)}{\partial t} + \frac{1}{r} \frac{\left(\partial r q_s\right)}{\partial r} = Q_s(r, t)
\end{align}
We have implemented QLKNN as a transport module inside JINTRAC and RAPTOR\cite{felici2018_raptorqlknn4d}. In the current implementation QLKNN provides the main ion heat flux $q_{i,1}$, electron heat flux $q_e$ and electron particle flux $\Gamma_e$. For multiple ion species we assume the same GyroBohm heat flux for each ion species. This involves a multiplication by ion density, and hence leads to negligible impurity heat flux for typical impurity densities. Contrary to RAPTOR which evolves electron density directly, JINTRAC solves the ion density equations for particle transport. Since the current version of QLKNN contains only $\Gamma_e$, we thus assume for each ion species $\Gamma_i = \frac{n_i}{n_eZ_i} \Gamma_e$, maintaining ambipolarity. This limits QLKNN, which is then not applicable for impurity transport or for multiple-isotope particle physics as in\cite{marin2019_multi_isotope}. The next generation of QLKNN will include multiple-ion transport. Finally, for numerical stability, we use either an effective diffusion $D_{eff}$ or convection $V_{eff}$, derived from the total particle flux, depending on the flux direction and density gradient. $V_{eff}$ is used for up-gradient particle transport and for low density gradients $\left(\left\lvert\frac{R}{L_n}\right\rvert<0.1\right)$. Future work will aim to improve on these assumptions by neural network fits on species dependent $D_i$ and $V_i$ directly, which is important for multiple-isotope fuelling and impurity transport applications.

Neural networks do not extrapolate well outside their training set boundaries. In this work, this is trivial to detect, as the training set was a bounded regular hyperrectangle. We chose to clip the inputs to the input layer of the neural network within the bounds of the hyperrectangle, with a margin of 5\% on all sides. Alternative approaches are also possible, such as training multiple neural networks to form a `committee', where extrapolation is detected from increased variance of the committee predictions in zones with sparse or non-existent data. This increase in variance arises from different local minima of the weight optimizations due to random initialisation. This is more suitable for training sets which are not pre-selected hyperrectangles, such as the training derived from experimental databases \cite{meneghini2017_core-pedestal-neural-network}. We chose not to implement this here due to the additional calculation times involved, and the trivial structure of our training set. Whether QLKNN remains within the bounds of the training set during application within integrated modelling can be found in post-processing.
\subsection{QLKNN simulation results within integrated modelling}
\label{sec:integrated_modelling}
We now compare QLKNN simulations to full QuaLiKiz within integrated modelling, for a representative set of 3 JET H-mode discharges. The correspondence between QuaLiKiz and the experimental profiles will not be discussed here, and on this point we refer the reader to the citations where the original JINTRAC-QuaLiKiz simulations were carried out for each of the cases. We focus on the correspondence between QuaLiKiz and QLKNN, as well as between the implementations within JINTRAC and RAPTOR.

To judge the quality of the neural network regression and the impact of the assumptions made, we show a comparison of QLKNN and QuaLiKiz on the high performance JET baseline \#92436\cite{ho2019_gp} within both JINTRAC and RAPTOR integrated modelling. The JINTRAC-RAPTOR comparison further acts as a benchmark exercise for correct coupling of QLKNN within the code suites. These simulations correspond to an averaged 500 ms time-window during discharge flattop. A Gaussian Process Regression fit is performed on the kinetic profile data, and the distribution average is used as initial condition\cite{ho2019_gp}. The current, temperature and density profiles are then evolved over multiple energy confinement times until the temperature and density profiles are in stationary-state, and compared to the experimental fits. As QLKNN is only applicable for turbulent transport in the tokamak core, we evolve temperature and density only inside $\rho_{N,tor} = 0.85$, and include a proxy transport coefficient for sawtooth-induced transport in the deep core for all simulations in this work.\cite{ho2019_gp}. Appendix \ref{sec:full_jetto_settings} contains a full overview of the applied settings. We show three simulations to investigate QLKNN model performance in different levels of increasing physics fidelity.

\begin{figure*}
    \includegraphics[width=\linewidth]{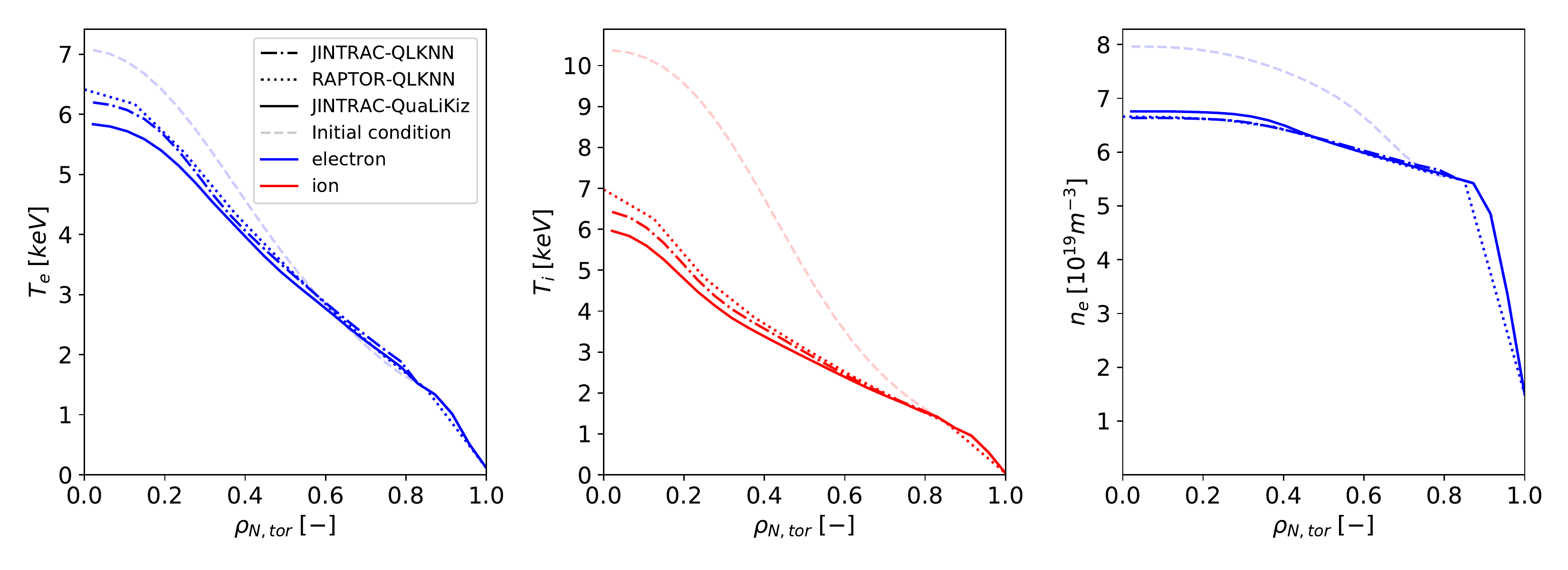}
  \caption{\label{fig:jintrac_raptor_92436}
  The final kinetic profiles of the JINTRAC-QuaLiKiz (solid) and JINTRAC-QLKNN (dash-dot) and RAPTOR-QLKNN(dot) simulations of JET shot \#92436. Shown are the final temperatures for the electrons (left, blue) and ions (middle, red) as well as the final electron density (right, blue). We also show the initial condition from GP regression in dashes. The simulations were done with low physics fidelity to aid the benchmark, most notably without rotation and with a pure plasma. The agreement between the three simulations is remarkable, within 5.5\% of each other. The QLKNN simulations were three orders of magnitude faster than the JINTRAC-QuaLiKiz one. Remaining small differences between JINTRAC and RAPTOR are likely caused by the different numerical schemes, and are currently under investigation.
  }
\end{figure*}
First we compare the implementation of QLKNN within JINTRAC and RAPTOR on a benchmark case. In both RAPTOR and JINTRAC we prescribe the ICRH and NBI power density calculated from the JET analysis chain, and a current density calculated from current diffusion integrated modelling with prescribed experimental kinetic profiles. A time-dependent equilibrium was calculated with ESCO within JINTRAC modelling. An equilbrium EQDSK file corresponding to the final time step was then generated and used within the JINTRAC-RAPTOR benchmark runs themselves, and kept constant during the runs. In JINTRAC we use a grid in $\rho_{N,tor}$ of 25 points. As RAPTOR uses a cubic spline base (order 3) compared to the finite differences scheme of JINTRAC (order 2), we need less points in RAPTOR to represent the same accuracy. As such, we use $25^{2/3} \approx 9$ points for $\rho_{N,tor}$. Both codes use the same boundary conditions at $\rho_{N,tor} = 0.85$ and the same initial condition, as taken from the GP fit of the experimental data. Neoclassical transport was not included. Finally, a pure plasma ($Z_{eff} = 1$) was assumed to reduce any effect of the different density equations being solved in RAPTOR (electrons) and JINTRAC (ions), and the radiation power loss was set to zero accordingly. We note two major implementation differences remaining between the codes. First, the spatial base on which the equations are solved, namely a sum of spatial basis functions (RAPTOR) and a finite difference scheme (JINTRAC), and the associated spatial smoothness. Secondly, RAPTOR uses a fully implicit scheme, leveraging analytical Jacobians of all equations, while JINTRAC is explicit using a dynamic time step and a predictor-corrector method to treat nonlinearities.

The final profiles of these simulations can be found in Figure \ref{fig:jintrac_raptor_92436}. The parameter of merit for QLKNN performance is the relative root mean square (RRMS) difference of the predicted kinetic profiles compared to the original QuaLiKiz runs. See Equation \ref{eq:rrms}, where the summation is over JINTRAC simulation radial grid points. The RRMS of the JINTRAC-QLKNN and RAPTOR-QLKNN runs are shown in Table \ref{tab:jetto_nn_92436_comp_results} and the final kinetic profiles and initial condition in Figure \ref{fig:jintrac_raptor_92436}.
\begin{table}
  \caption{\label{tab:jetto_nn_92436_comp_results} A comparison of the final kinetic profiles of JINTRAC-QuaLiKiz/QLKNN and RAPTOR-QLKNN benchmark case for JET \#92436. We use the relative root mean square profile difference (RRMS) in the region where QLKNN dominates transport, between the boundary condition $\rho_{BC}=0.85$ and deep core $\rho_{core}=0.212$. This simulation was done with lower physics fidelity to aid the benchmark, most notably a static equilibrium and a pure plasma without rotation. The differences between QLKNN and QuaLiKiz in JINTRAC are very small, and also match RAPTOR-QLKNN closely, all within 6\%. However, QLKNN was 3 orders of magnitude faster than the simulation with QLK.}
  \begin{ruledtabular}
    \begin{tabular}{*{1}{l} *{4}{c}}
      Comparison & \multicolumn{3}{c}{RRMS [\%]} &  \\
      & $T_i$ & $T_e$ & $n_e$ & $\rho_{core}$\\
      JINTRAC-QuaLiKiz vs JINTRAC-QLKNN & 3.6 & 5.1 & 1.2 & 0.212 \\
      RAPTOR-QLKNN vs JINTRAC-QLKNN & 2.4 & 4.2 & 0.7 & 0.212
    \end{tabular}
  \end{ruledtabular}
\end{table}
\begin{equation}
  \label{eq:rrms}
  RRMS \equiv \sqrt{\frac{\sum_{i=\rho_{BC}}^{\rho_{core}} (QLK_i - \hat{NN_i})^2}{\sum_{i=\rho_{BC}}^{\rho_{core}} QLK_i^2}}
\end{equation}
As is clear from the figure and RRMS, QLKNN performs remarkably well both in JINTRAC (maximum 5.1\% deviation from JINTRAC-QuaLiKiz) and in RAPTOR  (maximum 4.2\% deviation from JINTRAC-QLKNN). In both frameworks, this small hit in accuracy comes with momentous speed increases. The JINTRAC-QuaLiKiz run took 3.6 hours on 16 cores, while JINTRAC-QLKNN and RAPTOR-QLKNN took only 15 seconds and 8 seconds respectively on a single core, a speedup of three orders of magnitude. In JINTRAC, the evaluation of QLKNN itself was no longer the bottleneck, so only a modest speedup of \SI{2}{s} was obtained by parallelizing over 4 cores. Conversely, in RAPTOR the evaluation of QLKNN and its Jacobian matrix is the bottleneck, so further work will investigate utilizing MPI parallelization in the RAPTOR-QLKNN implementation, towards realtime simulation capability. Both simulations were run without using the MKL acceleration.

Two caveats have to be kept in mind when comparing the speeds of RAPTOR and JINTRAC. Firstly, the implicit scheme of RAPTOR allows for timesteps far exceeding transport timescales without numerical stability issues. However, to be able to capture transients we have set the timestep for RAPTOR to \SI{0.1}{s}. The Predictor-Corrector JINTRAC scheme is prone to numerical instabilities when the timestep becomes too large. For QuaLiKiz, this resulted in a maximum timestep of \SI{1e-3}{s}, which we also took for the JINTRAC-QLKNN run. It is possible to increase the timestep of JINTRAC-QLKNN to \SI{6e-3}{s} without resulting in instability, possibly by the smoothness of QLKNN in the unstable turbulent region compared to QuaLiKiz, resulting in a runtime of 7 seconds on a single core. Secondly, while the need to evaluate the Jacobian matrix of QLKNN results on a longer evaluation time per timestep, this allows RAPTOR to compute derivatives of the final state (e.g. kinetic profiles) to machine inputs (e.g. the ICRH input power), as well as derivative of intermediate plasma states. These derivatives are invaluable in control and optimization tasks, and are relatively expensive to compute using finite difference methods. The intermediate state derivatives can be used to find time varying linearized models of the plasma dynamics.

\begin{figure}
  \begin{minipage}{0.49\linewidth}
    \begin{center}
      92436 Rotationless
      \vspace{-.28cm}
    \end{center}
  \includegraphics[width=\linewidth]{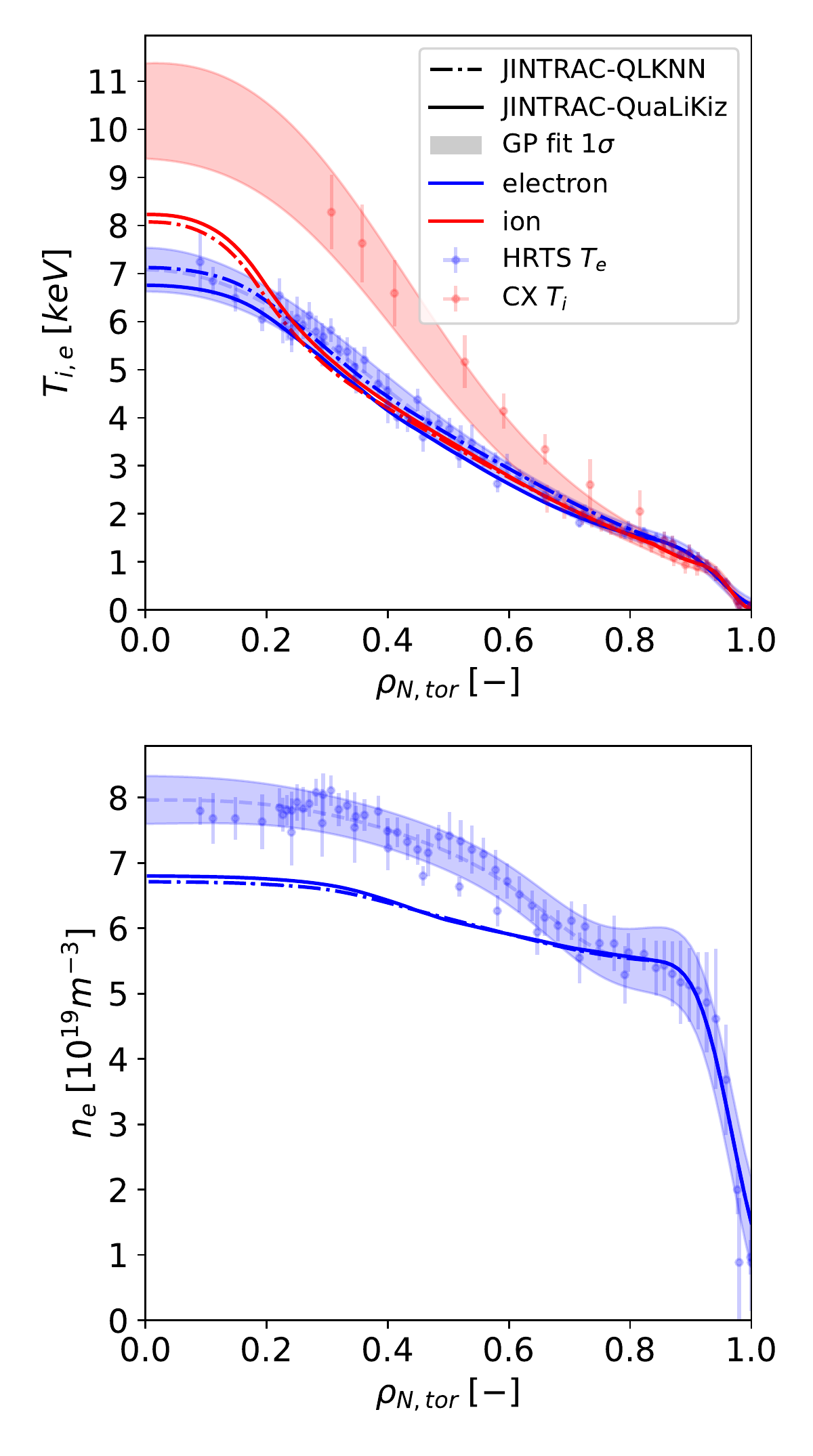}
  \end{minipage}
  \begin{minipage}{0.49\linewidth}
    \begin{center}
      92436 Full-physics
      \vspace{-.4cm}
    \end{center}
  \includegraphics[width=\linewidth]{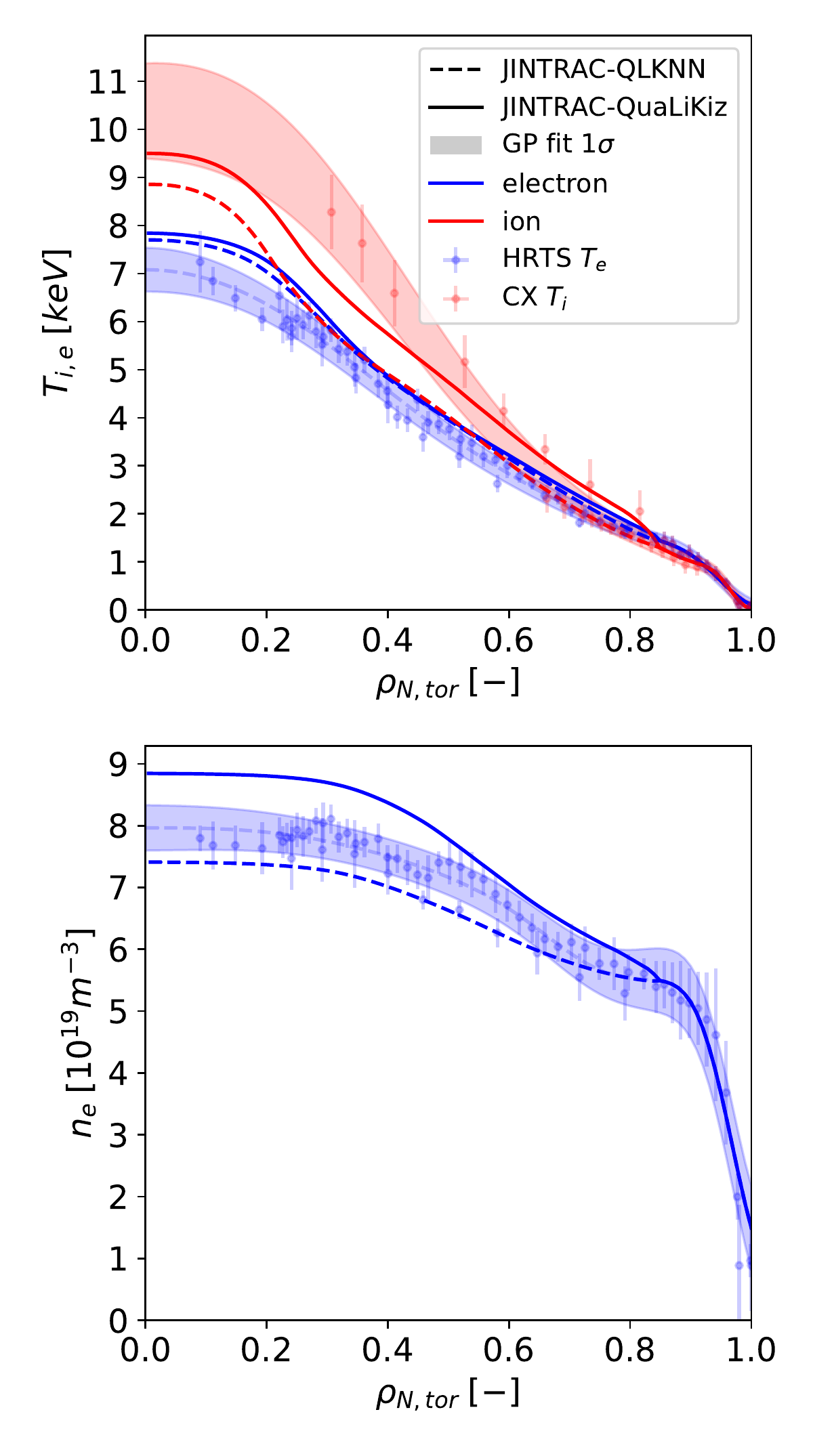}
  \end{minipage}
  \caption{\label{fig:nn_comp_92436} The final kinetic profiles of the JINTRAC-QuaLiKiz (solid) and JINTRAC-QLKNN (dash-dot) simulations of JET shot \#92436. Shown are the final temperatures for the ions (top, red) and electrons (top, blue) as well as the final electron density (bottom, blue). From left to right we show three cases of increasing physics fidelity: a reduced physics case, a more complete but rotationless case, and finally a case with rotation. Note the excellent agreement between QLKNN and QuaLiKiz in all figures, although a larger discrepancy was found for the case with rotation. This is expected, as the treatment of rotation is different in QuaLiKiz and QLKNN.}
\end{figure}

We then increase physics fidelity by using a more realistic mix of impurity isotopes, the inclusion of the ad-hoc electromagnetic stabilisation rule\cite{ho2019_gp}, and solving the magnetic equilibrium self-consistently with ESCO. These settings are described in detail in \cite{ho2019_gp} where the original JINTRAC-QuaLiKiz simulations were carried out. The final kinetic profiles are shown in the left column of Figure \ref{fig:nn_comp_92436}. In the right column we include the effect of rotation in the outer-half radius of the plasma\cite{citrin2017_qualikiz,ho2019_gp} using the QuaLiKiz native impact-of-rotation prediction for QuaLiKiz, and for QLKNN the new QLKNN-rotation-rule as described in section \ref{sec:rotation_rule}. The RRMS values are shown in Table \ref{tab:jetto_rest_nn_comp_results}.

Even in these simulations further from the QLKNN assumptions, the final kinetic profiles predicted by JINTRAC-QLKNN and JINTRAC-QuaLiKiz agree very well. The maximum discrepancy is on the order 1-15\%. The JINTRAC-QLKNN simulations were significantly faster, especially for the full-physics case, from around 7 days of walltime on 16 cores for JINTRAC-QLKNN to around 30 minutes of walltime on two cores for JINTRAC-QLKNN. This shows the applicability of QLKNN to reproduce approximately QuaLiKiz results for a fraction of the computational cost. The largest difference between QLKNN and QuaLiKiz can be found in the full-physics cases, as shown in Table \ref{tab:jetto_nn_92436_comp_results}. This discrepancy is mainly caused by the different treatment of rotation, as can be expected. While the inclusion of rotation did lead to an increase in $n_e$ and $T_i$ in the QLKNN simulations, for this case the degree of stabilisation is less than in QuaLiKiz itself.

Next we show the general applicability of QLKNN in two more JET shots. The first is the high collisionality baseline JET H-mode scenario \#73342\cite{baiocchi2015_cronosqlk,citrin_nn_proof_concept}, where the simulation corresponds to a stationary-state during flattop, and the GPR fit time-window was taken to be \SI{500}{ms}. The second case is high performance JET hybrid scenario \#92398, subject to DT extrapolation in upcoming campaigns\cite{casson2018_92398}. To demonstrate the capabilities of JINTRAC-QLKNN for dynamic evolution, this discharge was simulated during the density buildup following the LH transition. The GPR fits for each snapshot during the evolution was taken to be \SI{50}{ms}. Both cases were re-run with JINTRAC-QuaLiKiz for this paper, with interpretive impurities, meaning that the impurity profiles were constrained to match the main ion profile peaking, consistent with the QLKNN output assumptions.

\begin{figure}
  \begin{minipage}{0.49\linewidth}
    \begin{center}
      73342 Rotationless
      \vspace{-.28cm}
    \end{center}
  \includegraphics[width=\linewidth]{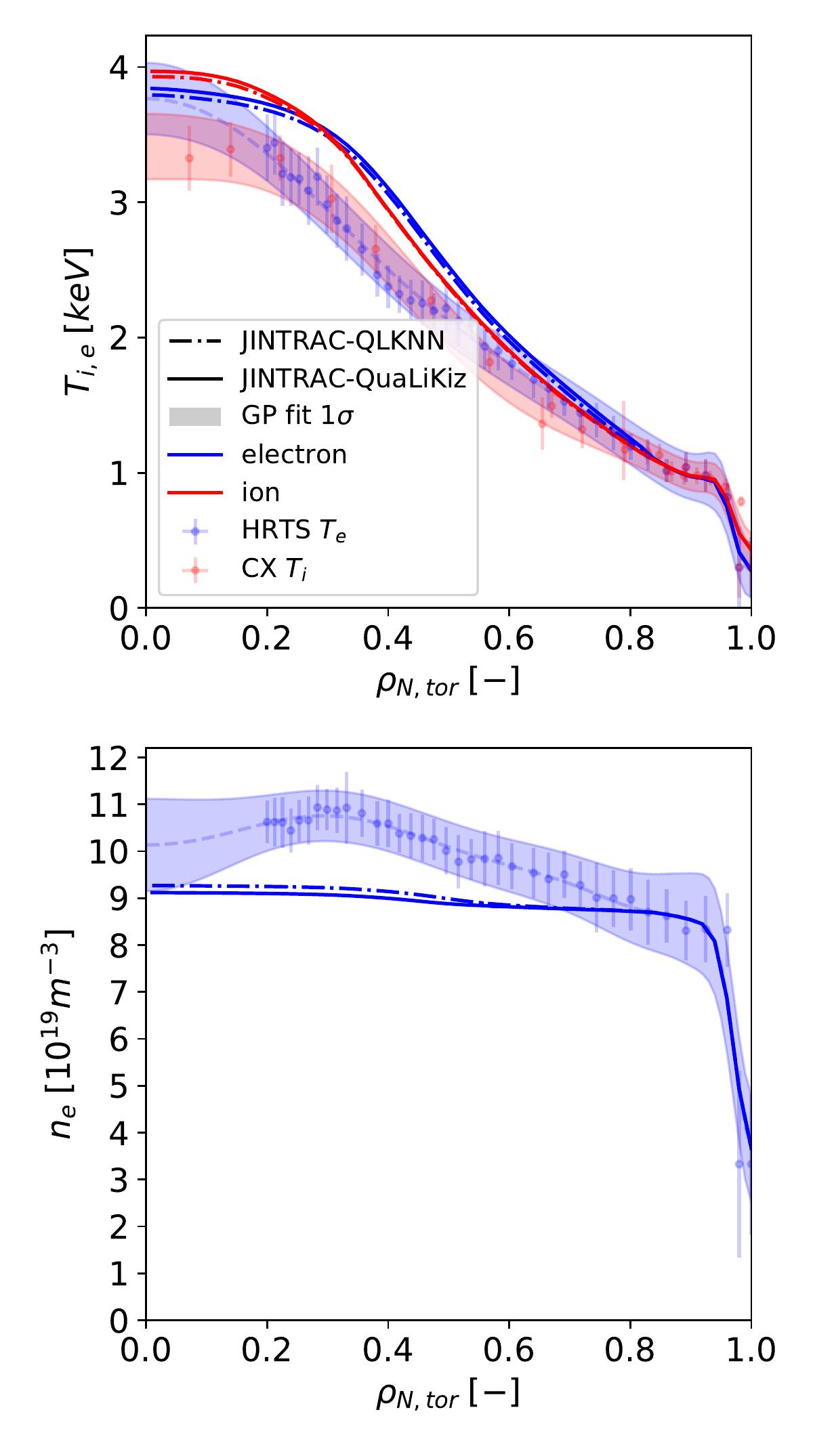}
  \end{minipage}
  \begin{minipage}{0.49\linewidth}
    \begin{center}
      73342 Full-physics
      \vspace{-.4cm}
    \end{center}
  \includegraphics[width=\linewidth]{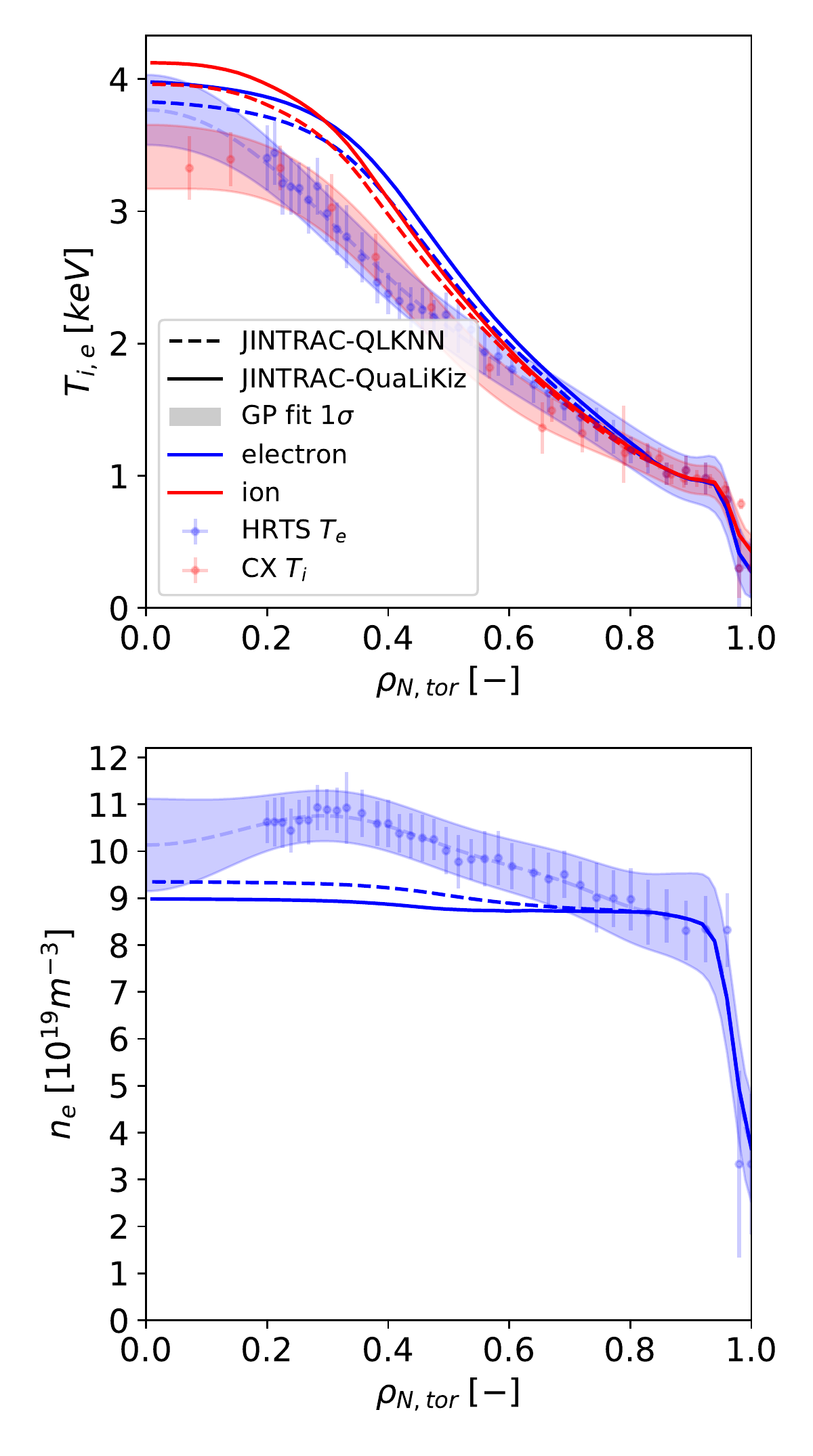}
  \end{minipage}
  \caption{\label{fig:nn_comp_73342}The final kinetic profiles of the JINTRAC-QuaLiKiz (solid) and JINTRAC-QLKNN (dash-dot) simulations of JET shot \#73342. Shown are the final temperatures for the ions (top, red) and electrons (top, blue) as well as the final electron density (bottom, blue). Both cases were run with interpretive impurities without rotation (left) and with rotation (right). The QLKNN predictions lie close to the QuaLiKiz ones, in the order of 4\% at maximum, which show the generality of applying QLKNN as quicker surrogate for the full QuaLiKiz model. We do note the underprediction of the density profile by QuaLiKiz. This is a known issue at high collisionality, which is currently under investigation by the QuaLiKiz team, and a revision of the collisionality model will be included in future releases.}
\end{figure}

\begin{figure}
  \begin{minipage}{0.49\linewidth}
    \begin{center}
      92398 Rotationless
      \vspace{-.28cm}
    \end{center}
  \includegraphics[width=\linewidth]{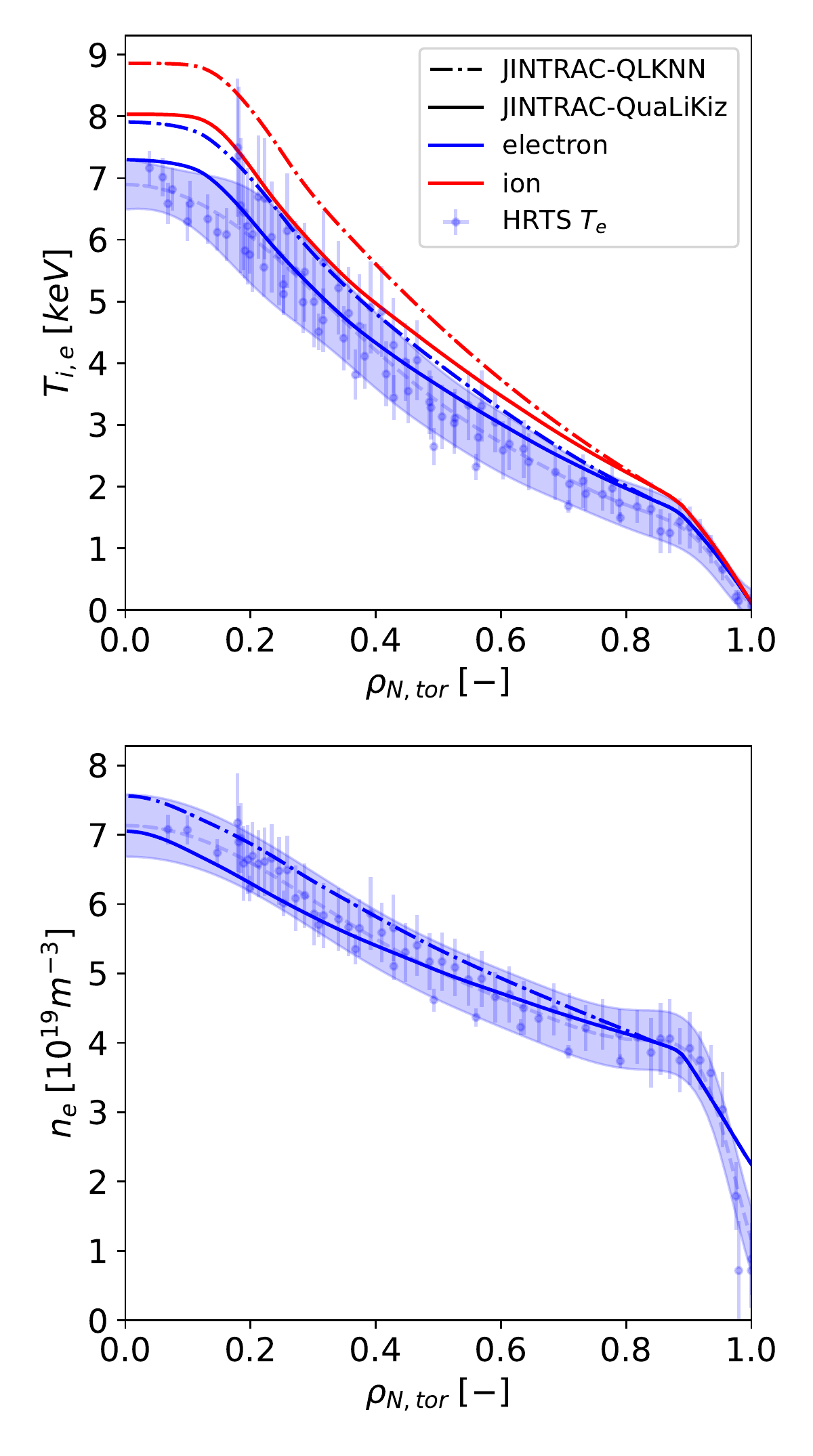}
  \end{minipage}
  \begin{minipage}{0.49\linewidth}
    \begin{center}
      92398 Full-physics
      \vspace{-.4cm}
    \end{center}
  \includegraphics[width=\linewidth]{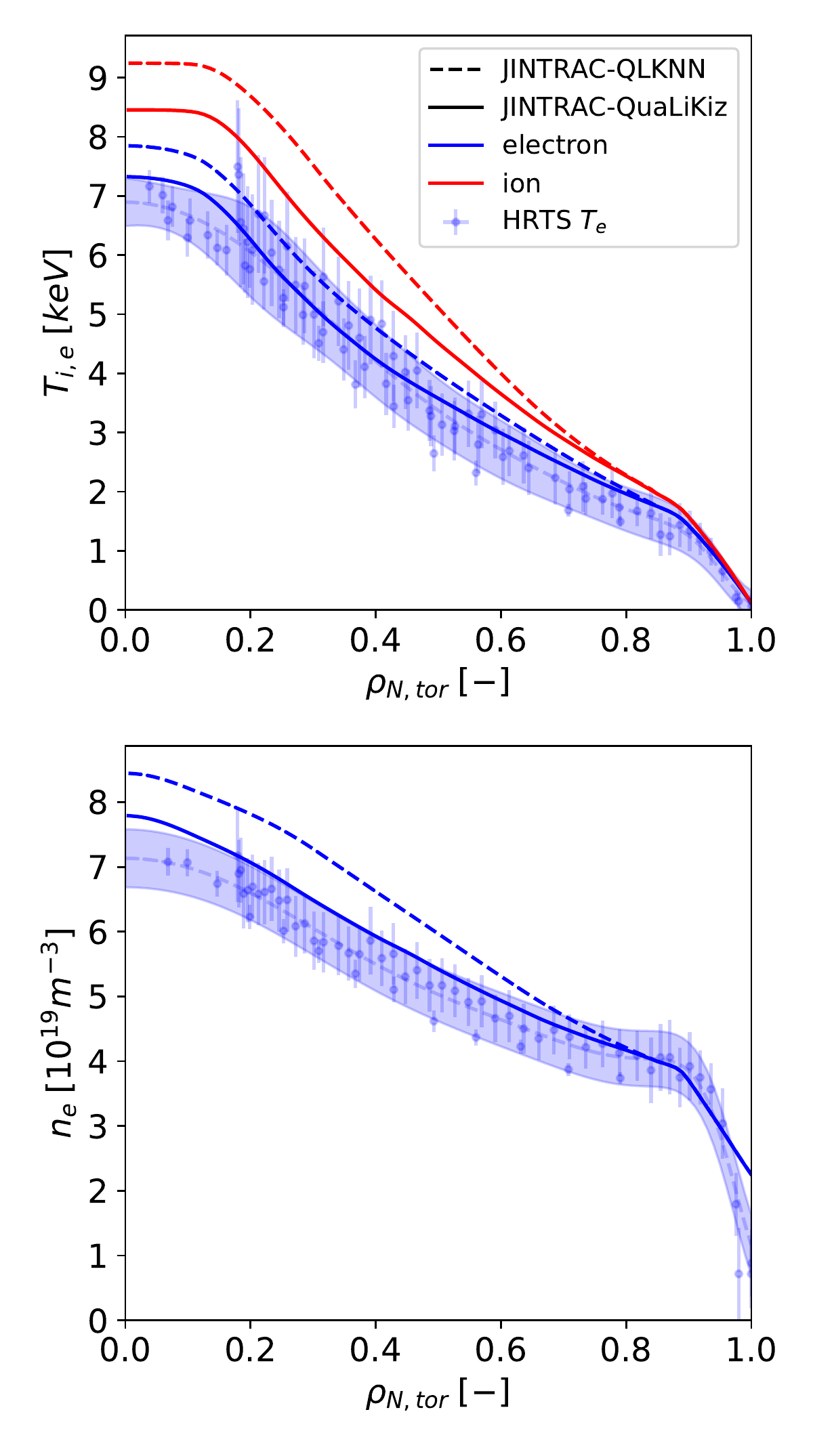}
  \end{minipage}
  \caption{\label{fig:nn_comp_92398}The final kinetic profiles of the JINTRAC-QuaLiKiz (solid) and JINTRAC-QLKNN (dash-dot) simulations of JET shot \#92398. Shown are the final temperatures for the ions (top, red) and electrons (top, blue) as well as the final electron density (bottom, blue). Both cases were run with interpretive impurities without rotation (left) and with rotation (right). Here the disagreement between QuaLiKiz and QLKNN is larger than previous cases, but still within 13\%. Future improvements to the QLKNN model are expected to lower these differences, but this result shows that even in this state the QLKNN model can be used for quick exploratory studies.}
\end{figure}

\begin{figure}
  \includegraphics[width=\linewidth]{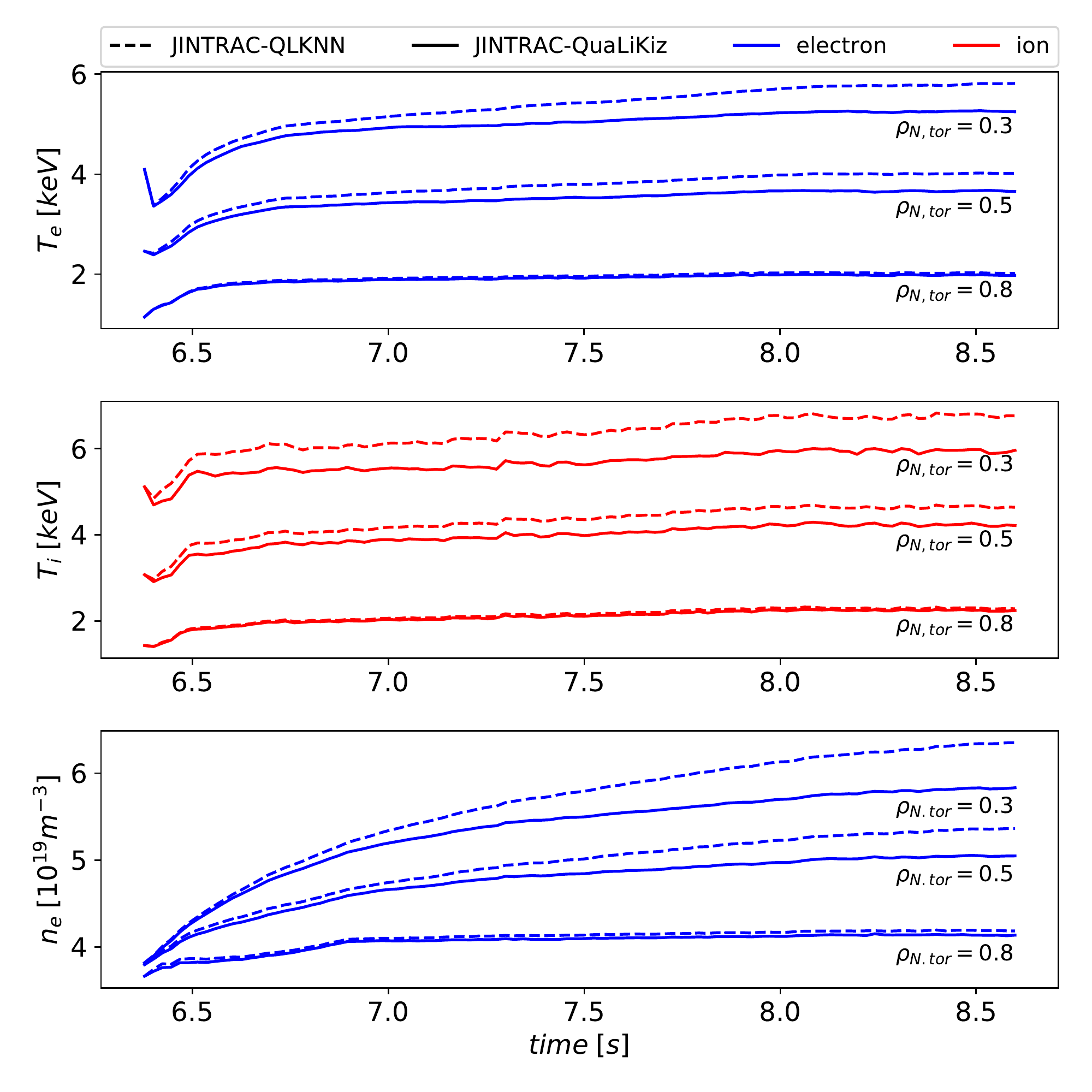}
  \caption{\label{fig:timedep}A time-dependent JINTRAC-QLKNN simulation without rotation of JET \#92398. Note the density buildup that is very well captured by QLKNN. The RRMS differences at $\rho=0.5$ for the full time-evolution are $T_e=8\%$, $T_i=9\%$, and $n_e=4\%$}
\end{figure}

The final kinetic profiles and comparison between QuaLiKiz and QLKNN for \#73342 are shown in Figure \ref{fig:nn_comp_73342} and Table \ref{tab:jetto_rest_nn_comp_results}, and for \#92398 in Figure \ref{fig:nn_comp_92398}. For \#92398 we also show the temperature and density temporal evolution at three radial locations in Figure \ref{fig:timedep}.

\begin{table}
  \caption{\label{tab:jetto_rest_nn_comp_results} The final kinetic profile differences between JINTRAC-QuaLiKiz and JINTRAC-QLKNN for the simulations of JET shot \#73342, \#92398 and \#92436. We use the relative root mean square profile difference (RRMS) in the region where QLKNN dominates transport, between the boundary condition $\rho_{BC}=0.85$ and core patch $\rho_{core}$. We show simulations without rotation (Rotationless), as well as with rotation (Full-physics). The differences between QLKNN and QuaLiKiz are small for the rotationless case, and larger for the full-physics case, mainly caused by the different treatment of rotation between QuaLiKiz and QLKNN.}
  \begin{ruledtabular}
    \begin{tabular}{*{1}{l} *{4}{c}}
      Simulation & \multicolumn{3}{c}{RRMS [\%]} & $\rho_{core}$  \\
      & $T_i$ & $T_e$ & $n_e$ &\\
      73342 rotationless & 0.5 & 1.6 & 1.1 & 0.25 \\
      73342 full-physics & 4.1 & 3.4 & 2.8 & 0.25 \\
      92398 rotationless & 12 & 10 & 7 & 0.15 \\
      92398 full-physics & 13 & 10 & 9.9 & 0.15 \\
      92436 rotationless & 3.1 & 7.5 & 0.7 & 0.212\\
      92436 full-physics & 2.8 & 15 & 14 & 0.212
    \end{tabular}
  \end{ruledtabular}
\end{table}
Again we note the excellent agreement between JINTRAC-QuaLiKiz and JINTRAC-QLKNN. The rotationless \#73342 case matches excellently within 2\%. \#92398 matches less well in comparison, around 10\%. While still good, we expect the match to improve by expanding the QLKNN input dimensionality, most notably a better capture of different impurity species and Shafranov shift, which are both planned in future work. \#73342 is with carbon wall (as per QLKNN training set assumptions) as opposed to ITER-like wall in \#92398, and \#92398 is high performance (high-$\beta$) and hence has more significant Shafranov shift.  We estimate that by expending 5-10 MCPUh it would be feasible to expand the range of the QuaLiKiz training set database sufficiently. Note that the better agreement between JINTRAC-QuaLiKiz and JINTRAC-QLKNN in the \#92436 case for the rotationless case compared to \#92398 may simply be coincidental, as the impact of the input dimensions not included in QLKNN can 'cancel out'.

For the cases with rotation, the impact on \#73342 is small, simply due to low rotation in this high-density case. For \#92398, the agreement between the native QuaLiKiz and QLKNN rotation rules is excellent, both boosting $n_e$ and $T_i$ significantly, and by the same magnitude. Note that $T_e$ is barely impacted by rotation, since the $T_e$ profile is predicted to be clamped by ETG turbulence for this discharge, both in the original QuaLiKiz and the QLKNN simulations.

The dynamic behaviour of QLKNN for \#92398 is shown in Figure \ref{fig:timedep}. The match between JINTRAC-QuaLiKiz and JINTRAC-QLKNN is excellent, most notably the density build-up in the lower plot, staying within a discrepancy of 4\% at mid-radius for the whole duration. However, the small differences between the two models compound from the outer-radius inward and over multiple timesteps, resulting in the relatively larger but still acceptable discrepancy for the final condition in Figure \ref{fig:nn_comp_92398}. While factor 4 less than compared to \#92436, the speed-up gained in the \#92398 simulation is still very significant, from 11 hours on 16 cores to 8 minutes on 2 cores.

%

\subsection{Physics-unaware network performance}
\label{sec:naive}
\begin{figure}
    \includegraphics[width=\linewidth]{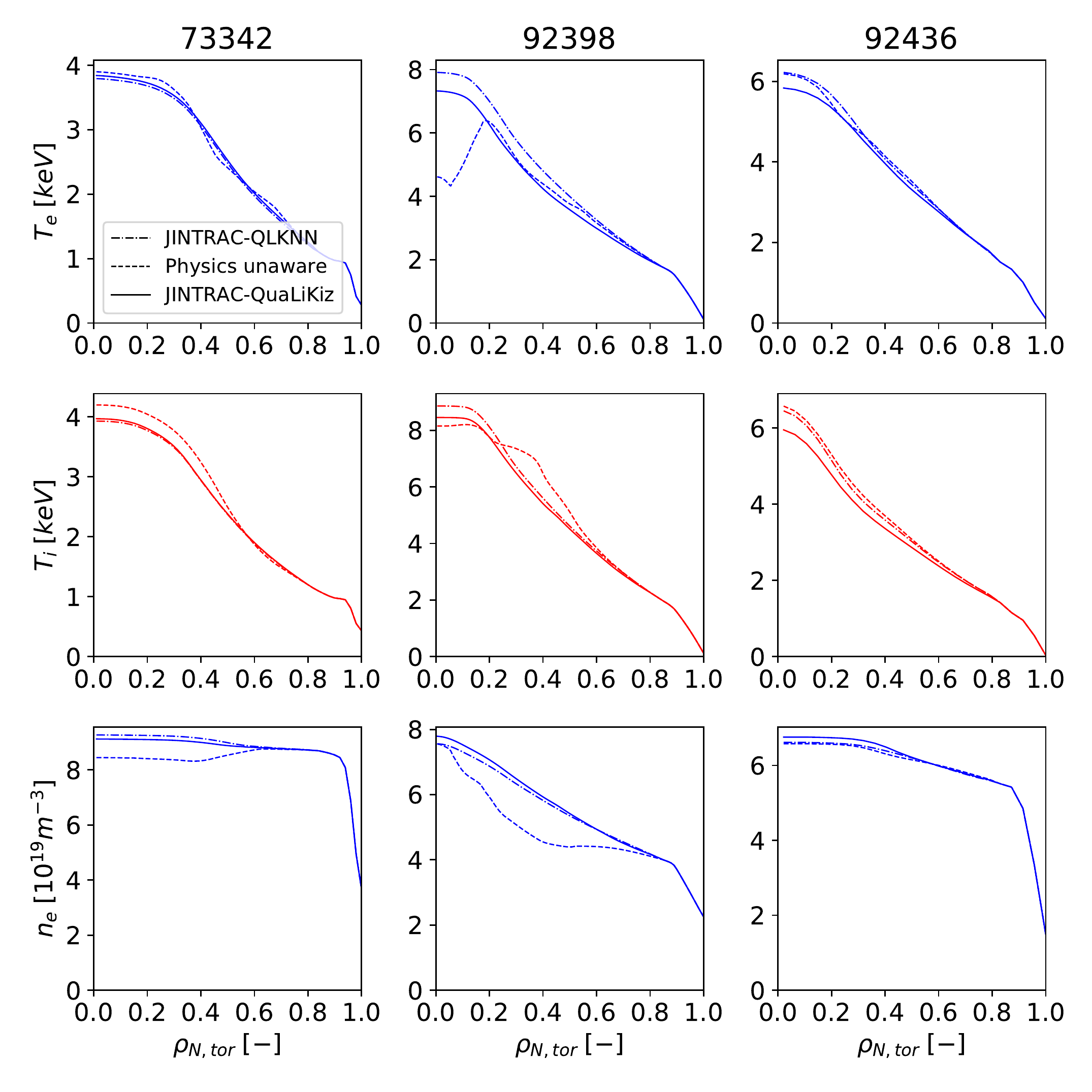}
  \caption{\label{fig:naive_net_jintrac} The final kinetic profiles of a JINTRAC simulation using a physics-unaware neural network model (dash-dot). The earlier shown JINTRAC-QuaLiKiz (solid) and JINTRAC-QLKNN (dashed) are shown for reference. As is clear, the physics-unaware model underperforms compared to JINTRAC-QLKNN in almost all cases.
  }
\end{figure}

Now that we have confirmed JINTRAC-QLKNN reproduces JINTRAC-QuaLiKiz well, we show results of a physics-unaware neural network model within JINTRAC. This model was trained on the same data as QLKNN, but was trained directly on the total $q_{i,e}$ and $\Gamma_e$ instead of on separate mode contributions, and did not employ the decomposition of fluxes to leading flux and flux ratios. Additionally, we used a standard machine learning cost function, as in \ref{eq:costfunction}. The final profiles of these physics-unaware network simulations can be found in Figure \ref{fig:naive_net_jintrac}.

Although there are cases where the physics-unaware networks perform (coincidentally) very similarly to QLKNN, notably for $T_i$ and $n_e$ in \#92436, clearly QLKNN performs better in almost all cases. The RMS errors on the full (stable + unstable) test set were not notably high, namely 1.1, 1.7 and 0.2 GB for $q_e$, $q_i$ and $\Gamma_e$ respectively. As such, we recommend looking beyond traditional measures of goodness when judging the quality of neural networks representing physical models.

\section{Conclusions and outlook}
We have shown a method to train physics-based neural networks as turbulent transport models, which we applied to generate a surrogate model for the fast quasilinear gyrokinetic transport model QuaLiKiz. Utilizing HPC, we generated a large dataset of $3\cdot10^8$ flux calculations, which was used as training set for fully connected feed forward neural networks for regression. Prior physics knowledge of the underlying model features was incorporated by using a customized cost function, choosing appropriate training targets, and looking beyond traditional measures of goodness. This surrogate model, QLKNN, has been integrated into two integrated modelling suites, JINTRAC and RAPTOR. We applied the JINTRAC-QLKNN ensemble to carry out predictive dynamic simulations of core transport in three JET shots, covering a representative spread of H-mode operating space. We have also shown one similar simulation using RAPTOR-QLKNN, in good agreement with JINTRAC-QLKNN. This benchmark was important for verifying the implementation of QLKNN in both code suites. The RAPTOR-QLKNN simulation calculation time was similar to JINTRAC-QLKNN in spite of the larger RAPTOR timesteps. This was due primarily due to the extra cost of calculating the derivatives of the output of the neural networks with respect to the input, as needed for the fully-implicit solver, as well as the extra calls needed for the implicit scheme Newton solver. However, these derivatives are invaluable in control and optimization tasks, and there is potential for further parallelization to bring down the evaluation time by a further order of magnitude towards realtime evaluation. The steady-state and dynamic kinetic profiles match those of the full QuaLiKiz simulations closely, while being up to five orders of magnitude faster to run.

The largest discrepancy between QLKNN and QuaLiKiz is caused by the different rotation rules employed between QLKNN and QuaLiKiz. The rotationless cases studied in this work showed differences from 1\%-10\% in the final kinetic profiles. The rotation cases studied showed mildly larger differences ranging from 3\%-15\%. The rotation discrepacy was more prevalent for the \#92436 case studied. An improved treatment of rotation will be part of future work, for example by implementing the quench rule on the individual growth rates in the spectrum before evaluating the saturation rule, thus capturing spectral shifts.

Future work will improve the QLKNN model by extending to larger input space, focusing on the impurity density gradient, and multiple-ion transport important for multiple-isotope fuelling applications and impurity transport. Additionally, using a robust method to fit a large amount of experimental kinetic profiles\cite{ho2019_gp}, one can base a training set on experimental data, instead of the hyperrectangle methodology described here. This allows for more input dimensions to be used, as well as including rotation by using the native QuaLiKiz rotation model, instead of a rotation rule as described here. There are also other techniques to include physics information in neural networks. The 'late fusion' method can be used to include functional information in the network architecture itself, for example by constraining the mapping to a critical gradient model, and has already been successfully used in a proof-of-principle QuaLiKiz surrogate model\cite{daniel2019_thesis}. Finally, instead of fitting the transport fluxes directly with a neural network, more primitive linear characteristics can be fit for the entire spectrum, e.g. growth rates, frequencies, phase-shifts. The transport flux calculations would then arise from application of a nonlinear saturation rule in post-processing of neural network outputs, which is a trivially fast calculation, allowing the rapid testing and evaluation of multiple saturation rules.

Beyond the model improvements, work can now commence on extensive experimental validation of QLKNN predictions, as well as using QLKNN for scenario optimisation and design. As shown in this work, physics-based neural network surrogate models can enable first-principle dynamic transport simulations at unprecedented speeds, opening up new avenues for tokamak scenario optimization and realtime control applications.
\begin{acknowledgments}
This work has been carried out within the framework of the EUROfusion Consortium and has received funding from the Euratom research and training programme 2014-2018 and 2019-2020 under grant agreement No 633053. The views and opinions expressed herein do not necessarily reflect those of the European Commission.
This research used resources of the National Energy Research Scientific Computing Center (NERSC), a U.S. Department of Energy Office of Science User Facility operated under Contract No. DE-AC02-05CH11231.
We are grateful to F. Jenko for assistance with computational resources. We would also like to thank the open source community for developing all these great tools, most notably NumFOCUS for supporting scientific tools as numpy, pandas and xarray. Also the Bokeh team for their excellent plotting tools, and the peewee team for making databases accessible for the beginner.
\end{acknowledgments}
\section*{Data provenance}
All data in this manuscript is stored according to the FAIR principles. Plots, underlying data, and metadata is stored by Zenodo, and can be found at \href{https://doi.org/10.5281/zenodo.3595558}{10.5281/zenodo.3595558}.
%
\newpage
\section*{References}
\bibliography{Bibliography}
\appendix
\section{Per-mode transport flux contribution calculation}
\label{sec:mode_split}
To aid with successful neural network regression, as discussed in the subsequent sections, QuaLiKiz was modified to additionally output fluxes and transport coefficients arising solely from individual classes of modes, i.e. ITG, TEM, ETG. Mode identification is determined by mode number (ion or electron scale) and mode frequency (ion or electron direction). The ETG electron heat flux is defined as the $q_e$ arising from the spectrum $k_\theta\rho_s>2$. To separate ITG and TEM fluxes, the saturation rule was evaluated twice at ion-scales, for electron modes and ion modes separately. This is different from the regular QuaLiKiz scheme, where the saturation rule is evaluated once for all modes at ion-scales. This can lead to inconsistencies comparing regular QuaLiKiz and these newly created transport flux and transport coefficient outputs. In other words, for saturation rule $\text{SAT}$ and ITG spectrum $k_{ITG}$ and TEM spectrum $k_{TEM}$, $\text{SAT}(k_{ITG} \cup k_{TEM}) \neq \text{SAT}(k_{ITG}) + \text{SAT}(k_{TEM})$. However, in practice, the difference between summing the separate ITG and TEM fluxes together (in cases where they coexist in the spectrum) compared to their self-consistent total evaluation in the saturation rule, is typically less than $20\%$. To further extend the general applicability of the neural networks, we use a form of GyroBohm normalization for all transport coefficients in this work, as defined in Appendix \ref{sec:gyrobohm}.
\section{QLKNN GyroBohm normalizations}
We define here the GyroBohm normalizations used by QLKNN. First, the normalizations of the predicted transport fluxes:
\label{sec:gyrobohm}
\begin{align}
  \label{eq:gyrobohm1}
  c_{GB} &\equiv \frac{\sqrt{A_{i,0} m_p}T_{e,SI}^{1.5}}{q_e^2 B_0^2 a} \\
  \Gamma_{s,GB} &\equiv \frac{a}{n_s c_{GB}}\Gamma_{s,SI} \\
  D_{s,GB} &\equiv \frac{1}{c_{GB}}D_{s,SI} \\
  V_{s,GB} &\equiv \frac{a}{c_{GB}}V_{s,SI} \\
  q_{s,GB} &\equiv \frac{a}{n_s T_{s,SI} c_{GB}}q_{s,SI} \\
  \chi_{s,GB} &\equiv \frac{1}{c_{GB}}\chi_{s,SI}
  \label{eq:gyrobohm2}
\end{align}
The transport coeffients are denoted with SI for fluxes in SI units, and GB for fluxes in GyroBohm units. $a$ and $R$ are the midplane-averaged minor and major radii of the last-closed-flux-surface. Furthermore, $q_e$ is the electron charge, $B_0$ is the magnetic field at the magnetic axis, $m_p$ the proton mass, and $A_{i,0}$ the atomic number of the main ion. Unless noted otherwise, all radial derivatives are against the midplane-averaged minor radius $r \equiv r_{minor}$. For convenience, we define the normalized length scales:

\begin{align}
  L_{T_{i,e}} \equiv -T_{i,e} \left(\frac{\dif T_{i,e}}{\dif r}\right)^{-1} \\
  L_{n} \equiv -n \left(\frac{\dif n}{\dif r}\right)^{-1}
\end{align}
the normalized collision frequency:
\begin{align}
  \nu^* &\equiv \nu^*_e \equiv \nu_e\tau_{bounce} \\
  \nu_e &\equiv 917.4 Z_{eff} (10^{-19} n_e) \Lambda_e (10^{3} T_e)^{-1.5} \\
  \Lambda_e &\equiv15.2-0.5 \ln(10^{-20}n_e)+\ln(10^{3} T_e) \\
  \tau_{bounce} &\equiv\frac{q R}{{(\frac{r}{R})^{1.5} \sqrt{\frac{q_e}{m_e} 10^{3} T_e}}}
\end{align}
where $q_e$ and $m_e$ are the electron charge and mass respectively, and finally the effective ion charge $Z_{eff}$:
\begin{equation}
  Z_{eff} \equiv \frac{\sum_i n_i Z_i^2}{\sum_i n_i Z_i} = \frac{\sum_i n_i Z_i^2}{n_e}
\end{equation}
\section{Introduction to neural networks}
\label{sec:neural_networks}
\begin{figure}[h]
  \begin{minipage}{0.49\linewidth}
    \centering
    \begin{equation*}
      y = f \left( \sum_{i=1}^n w_i x_i + b \right)
    \end{equation*}
  \end{minipage}
  \begin{minipage}{0.49\linewidth}
    \centering
    \begin{tikzpicture}[shorten >=1pt,->,draw=black!50, node distance=\layersep]
      \node[hidden neuron, pin={[pin edge={<-}]180-35:$x_0$}, pin={[pin edge={<-}]180:$x_i$}, pin={[pin edge={<-}]180+35:$x_n$}, pin={[pin edge={->}]right:$y$}] (something) {};
    \end{tikzpicture}
  \end{minipage}
  \caption{\label{fig:single_neuron}Schematic and mathematical representation of a neuron. A neuron takes multiple inputs, multiplies each input with a weight $w_i$, sums the results, adds a bias $b$ and applies an activation function $f$. Generally $f$ is a nonlinear function.}
\end{figure}
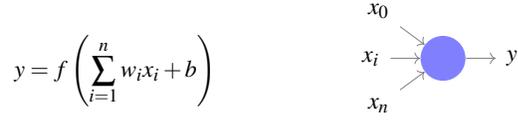

Neural networks are universal approximators and hence a powerful tool for regression \cite{nn_foundation}. In this work we apply fully connected feed-forward neural networks to a supervised regression problem, in which we reproduce the input-output mapping of the QuaLiKiz code. The basic building block of a FFNN is the \textit{neuron}, with \textit{activation function} $f$, as shown in figure \ref{fig:single_neuron}.

In a FFNN, neurons are distributed into layers, with each neuron in a layer taking the output of each neuron in the previous layer as input. Most FFNNs have at least an \textit{input layer} in this case taking the physical input \textit{features} described in Section \ref{sec:dataset}, a \textit{hidden layer} capturing the to-be learned hidden relationships and an \text{output layer}, combining the learned relationships into a \textit{target}. A FFNN with a single hidden layer is able to reproduce any sufficiently smooth input-output mapping up to arbitrary error \cite{cybenko1989_universial_approximation}, but in practice training a network with at least two hidden layers has better regression and convergence properties.

Using the notation of Ref. \cite{nndl_book}, namely $l$ as the number of the current layer, $j$ for the number of the neuron in the current layer and $k$ the number of the neuron in the previous layer. A neuron is then fully defined by its weight $w$, bias $b$, activation function $f$ and output or \textit{activation} $a$, see Figure \ref{fig:ffnn_example}, or the one-neuron equation \ref{eq:one_neuron}:
\begin{equation}
  \label{eq:one_neuron}
  a^l_j = f \left( \sum_k w^l_{jk} a^{l-1}_k + b^l_j \right)
\end{equation}
These layers can be arbitrarily combined. For example we show the explicit formula for a 2-hidden layer neural network with N-dimensional input $x_{in}$ and M-dimensional output $y$ in Equation \ref{eq:ffnn_example}. The output layer has a linear activation function which is simply the identity function $f \left(x\right) = I(x) = x$, as is usual for regression problems. We also assume each hidden layer has the same nonlinear activation function $\sigma$.

\begin{equation}
  \label{eq:ffnn_example}
  y = a^3_1 = \sum_i^M w^{3}_{i} \sigma \left(\sum_j^O w^2_{ij} \sigma \left( \sum_k^N w^1_{jk} x_{in,k} + b^1_j \right) + b^2_i \right) + b^{3}_o
\end{equation}
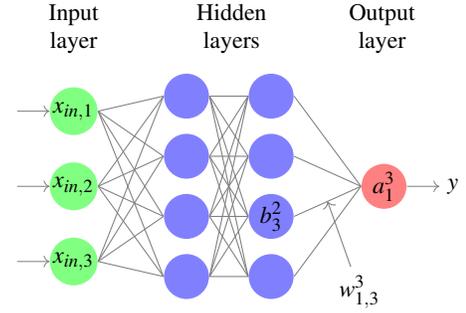
\begin{figure}[h]
  \begin{tikzpicture}[-,draw=black!50, node distance=\layersep cm]
    \def\layersep{1.5}
    \def\prefix{NN1}
    \def\inputnodes{3}
    \def\hiddenlayers{2}
    \def\hiddennodeblocks{4}
    \def\outputnodes{1}
    \def\outputlabel{$y$}
    \pgfmathparse{(1+0.75*(\hiddenlayers-1))*\layersep}\let\lastlayerxcoord=\pgfmathresult
    \pgfmathparse{-(\hiddennodeblocks * \nodedist)}\let\lastnodeblockycoord=\pgfmathresult
    \pgfmathparse{\hiddennodeblocks+1}\let\secondblockstart=\pgfmathresult
    \let\secondblockend=\hiddennodeblocks
    \foreach \name / \y in {1,...,\inputnodes}
    \node[input neuron,  pin={[pin edge={<-}]180:}] (\prefix-I-\name) at (0,-\y) {$x_{in,\y}$};

    \foreach \layernum in {1,...,\hiddenlayers} {
      \pgfmathparse{(1+0.75*(\layernum-1))*\layersep}\let\layerxcoord=\pgfmathresult
      \foreach \yy in {1,...,\hiddennodeblocks} {
        \pgfmathparse{-(\yy * \nodedist)}\let\nodeycoord=\pgfmathresult
        \ifthenelse{\equal{\yy}{3}\and\equal{\layernum}{2}} {
          \node[hidden neuron] (\prefix-H\layernum-\yy) at (\layerxcoord,\nodeycoord) {$b^2_3$};
        } {
          \node[hidden neuron] (\prefix-H\layernum-\yy) at (\layerxcoord,\nodeycoord) {};
        }
      }
    }

    \node[output neuron,pin={[pin edge={->}]right:\outputlabel}] (\prefix-O) at (\lastlayerxcoord+\layersep, \lastnodeblockycoord+1.2) {$a^3_1$};

    \foreach \source [evaluate=\source using int(\source)] in {1,...,\inputnodes}
    \foreach \dest[evaluate=\dest using int(\dest)] in {1,...,\secondblockend}
    \path (\prefix-I-\source.east) edge (\prefix-H1-\dest.west);

    \foreach \source [evaluate=\source using int(\source)] in {1,...,\secondblockend}
    \foreach \dest[evaluate=\dest using int(\dest)] in {1,...,\secondblockend}
    \foreach \layernum [remember=\layernum as \prevlayernum (initially 1)] in {2,...,\hiddenlayers}
    \path (\prefix-H\prevlayernum-\source.east) edge (\prefix-H\layernum-\dest.west);

    \foreach \source [evaluate=\source using int(\source)] in {1,...,\secondblockend}
    \path (\prefix-H\hiddenlayers-\source.east) edge (\prefix-O.west);

    \node[] (weight) at (3.8,-3.4) {$w^3_{1,3}$};
    \draw[->] (weight) -- ($ (NN1-H2-3) !.5! (NN1-O) $);


    \node[annot] (h1) at (0, .1) {Input layer};
    \node[annot] (h2) at (2.1,.1) {Hidden layers};
    \node[annot] (h2) at (4.1,.1) {Output layer};
\end{tikzpicture}
\caption{\label{fig:ffnn_example} A schematic representation of a two-hidden-layer feed-forward neural network. $w^l_{jk}$ is the weight of the connection from the $k\mathth$ neuron in the $(l-1)\mathth$ layer to the $j\mathth$ neuron in the $l\mathth$ layer. Then, $b^l_j$ is the bias of the $j\mathth$ neuron in the $l\mathth$ layer. The final network output is the activation of the $j\mathth$ neuron in the $l\mathth$ layer.}
\end{figure}

The weights and biases of the network are determined by minimizing some cost function or \textit{loss function} $C$, called \textit{training}. Assuming we have $S$ input-output mapping \textit{samples}, we collect these in an $N \times S$ \textit{input matrix} and $M \times S$ \textit{output matrix}. Before training the full dataset is generally split in a \textit{training set}, which is used to update the weights and biases, a \textit{validation set} which is used to check generalization of the neural network model after every step of the optimizer algorithm, and a \textit{test set} which is not used at all during training, and is used to check generalization across any tunable parameters related to the training process described later. The weights and biases are updated using an optimizer, usually a variant of (mini-batch) gradient descent\cite{bottou2016_optimization_gradient_descend}. For mini-batch gradient descent the training set is further split in batches of size $B$, which is itself an hyperparameter to be tweaked. A small batch size will generally be slower to converge, as some vectorization of internal calculations is lost, but the resulting model has a better generalizing properties\cite{masters2018_batch_size}. A common choice for measure-of-goodness and regularizing term for regression tasks are the mean square error and L2-regularization respectively. Now we can write down a general formula for the cost function, extended in this work in Section \ref{sec:cost_function}, where $y_i$ is the network prediction for a single sample, and $\hat{y}_i$ the real value in the dataset:

\begin{align}
\label{eq:costfunction}
  C = C_{good} + \lambda_{regu} C_{regu} \\
  C = \frac{1}{B} \sum^B_i (\hat{y}_i - y_i)^2 + \lambda_{regu} \lVert W \rVert_2^2
\end{align}
Where $\lVert W \rVert_2$ denotes the matrix L2-norm of all the weights combined. The derivative of the cost function with respect to its tunable parameters can be analytically determined using the chain-rule in what is called \textit{backpropagation}. This can then be used in the update of the gradient descent, generally of the form:

\begin{align}
  \theta_{n+1} = \theta_n - \gamma \nabla C(\theta_n)
\end{align}
where $\theta$ are the tunable parameters ($w$ and $b$) and $\gamma$ is the step size or \textit{learning rate}, another hyperparameter to be optimized. The training algorithm needs an initial guess $\theta_0$ to start training, which is in our case a random Gaussian distribution with mean $0$ and standard deviation $1$ for all weights and biases. The weights and biases are updated every batch $B$. After the optimizer has seen the full training set, i.e. all batches, this is called an \textit{epoch}. The resulting neural network is then used to determine the loss against the full validation set, which is used to determine convergence. If convergence is reached, the training is stopped and the neural network saved. If not, all samples are re-shuffled and new batches are taken, repeating this procedure until convergence is reached. In this work we use \textit{early stopping} to determine convergence. Early stopping sets a bound on the amount of epochs the loss of the validation set is allowed to increase, a hyperparameter called \textit{patience}. Early stopping prevents overfitting and gives a robust stopping criterion.

This method of training is quick, even for a large amount of parameters, as $\nabla C(\theta_n)$ is analytical and efficient to calculate. It is thus also quick to calculate the derivatives of the final trained neural network with respect to its inputs $\dif y / \dif x$. This is highly useful for our application, when the neural network turbulent surrogate models are integrated into implicit PDE solvers (solving the transport equations), used for trajectory control applications, or applied to tokamak scenario optimization.
\section{Physics-based training targets}
\label{sec:training_targets}
\begin{figure} 
  \begin{tikzpicture}[-,draw=black!50, node distance=\layersep cm, every node/.style={scale=0.6}]
    \path (0,3) pic {ffnn={NN2 inputnodes 4 hiddenlayers 2 hiddennodeblocks 2 outputnodes 1 outputlabel $\Gamma_e/q_e$ twoblock 1}};
    \path (.75,0) pic {ffnn={NN1 inputnodes 4 hiddenlayers 2 hiddennodeblocks 2 outputnodes 1 outputlabel $q_e$ twoblock 1}};
    \path (0,-3) pic {ffnn={NN3 inputnodes 4 hiddenlayers 2 hiddennodeblocks 2 outputnodes 1 outputlabel $q_i/q_e$ twoblock 1}};
    \node[output neuron,pin={[pin edge={->}]right:$\Gamma_e$}] (O) at (5,0) {};
    \path (NN1-O) edge (O);
    \path (NN2-O) edge (O);

    \node[output neuron,pin={[pin edge={->}]right:$q_i$}] (O2) at (5,-3) {};
    \path (NN3-O) edge (O2);
    \path (NN1-O) edge (O2);
  \end{tikzpicture}
  \caption{Schematic overview of the $\Gamma_e$ and $q_e$ predicted by a combined leading flux and ratio-predicting neural network for TEM fluxes. Three separate FFNNs, one predicting the leading flux $q_e$ and two ratio-predicting network predicting $\Gamma_e/q_e$ and $q_i/q_e$ are combined to a network ensemble that predicts $\Gamma_e$, $q_e$, and $q_i$.}
  \label{fig:combo_net}
\end{figure}
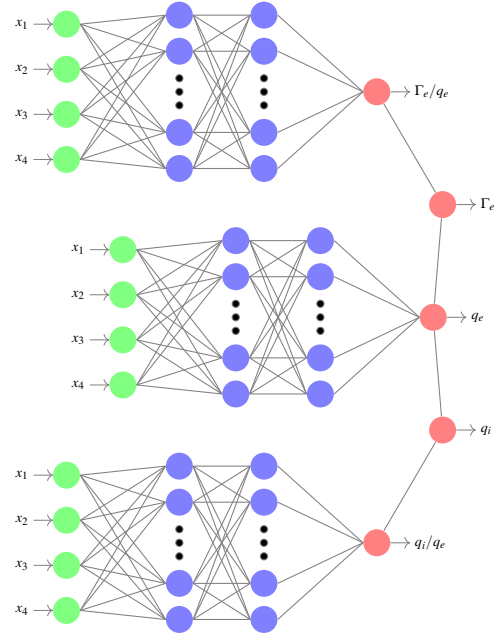

The identical critical thresholds for all transport channels was forced by a careful choice of training targets. The transport coefficients were separated into a \textit{leading flux} and \textit{flux ratios}. For example, for TEM fluxes, the leading flux is the electron heat flux $q_e$, resulting in the flux ratios $q_i/q_e$ and $\Gamma_e/q_e$. Networks are then trained on the leading flux and flux ratios separately, resulting in a leading flux network, and flux ratio networks. In the transport model implementation, the flux ratio predictions and leading flux predictions are multiplied together to re-obtain the original transport fluxes $q_i$ and $\Gamma_e$. This procedure is sketched in Figure \ref{fig:combo_net}. The fact that the leading flux is zero in the stable region (below the critical threshold), guarantees that the thresholds of all transport channels are identical. Increased smoothness and quality in the regression is achieved by removing training set outliers through data filtering (see section \ref{sec:filtering}). The mode separation technique works well with QuaLiKiz, where the ITG, TEM, and ETG modes are always clearly separable and identifiable. This technique may be more challenging in higher-fidelity models, or in future QuaLiKiz extensions, where it is uncertain if multiple additional co-existing modes (e.g. micro-tearing-modes, kinetic ballooning modes, fast-ion driven modes) are always clearly identifiable and separable.

Splitting the training targets by mode (ITG, TEM, ETG) and training a single network for each was found important for obtaining flux ratio regressions of sufficient quality. Letting a single network predict multiple channels was found to introduce cross-talk between the channels, resulting in the same averaged total error, but a different error per channel. For example, depending on the random initialization, ion heat flux would be better than average and the electron heat flux worse than average. Then, for the next network, with different initialization, the electron heat flux would be better than average and vice versa. As no averaging between networks was used in this work, we chose to separate the transport channels in separate networks.

\begin{figure} 
  \includegraphics[width=\linewidth]{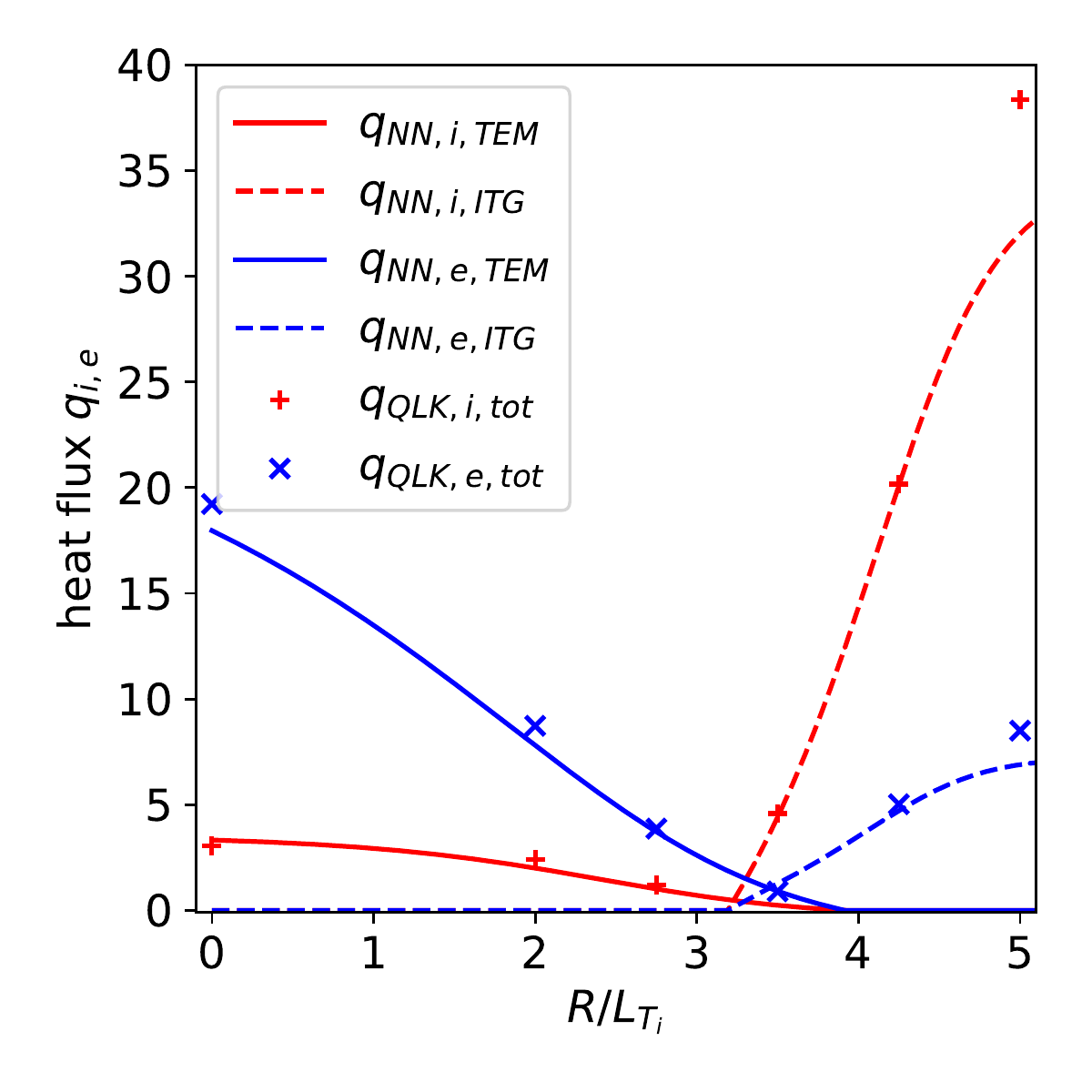}
  \caption{\label{fig:tem_itg}The QuaLiKiz predicted total heat fluxes for electrons $q_{QLK,e,tot}$ and ions $q_{QLK,i,tot}$ for multiple values of $\frac{R}{L_{T_i}}$, while keeping the other input parameters constant (pluses and crosses). We also show neural networks fit with the methodology described in \ref{sec:physics_based_nn}, denoted with $NN$ and $TEM$/$ITG$ for networks trained on the separate TEM and ITG fluxes respectively. Important to note is the capture of sharp transport characteristics around $-\frac{R}{T_i}\frac{dT_i}{dr} \approx 3.1$. Note the excellent quality of regression throughout. The discrepancy for the highest $\frac{R}{L_{T_i}}$ point is due to it being filtered out of the training set due to non-experimentally-relevant high flux, see section~\ref{sec:filtering}}
\end{figure}

Flux ratio network training for total fluxes (i.e., corresponding to the original QuaLiKiz output, as opposed to each of the separated ITG, TEM, ETG flux outputs) was unable to converge to a result of sufficient quality for a robust surrogate turbulence model, even after extensive hyperparameter scans. This is likely due to sharp discontinuities present in the flux ratios when not separating the fluxes. This is apparent in a TEM-ITG transition, for example in a scan of $R/L_{T_i}$ as shown in Figure \ref{fig:tem_itg}. The boundary between ITG and TEM regimes for this specific parameter set is $-\frac{R}{T_i}\frac{dT_i}{dr} \approx 3.1$. Above this value (ITG regime), $q_i/q_e>1$. Below this value (TEM regime), $q_i/q_e\ll1$. The transition between these regimes is extremely sharp, a feature challenging to capture by a regularized neural network. Instead, we use the mode-specific fluxes calculated by QuaLiKiz described in Section \ref{sec:dataset}, where the mode-specific flux ratios within the separate ITG and TEM regimes are more uniform compared to the total flux ratio. The output of the per-mode predicting networks are then added together in the transport model implementation in postprocessing using an unweighted sum. This results in a small difference between QuaLiKiz predicted fluxes and QLKNN predicted fluxes in regions where ITG and TEM coexist, as mentioned in Section \ref{sec:dataset}. Fitting the separate modes results in clearer thresholds without transition regions, enabling the use of the modified cost function in \ref{sec:cost_function}, resulting in a sharper transition at the threshold. As seen in figure \ref{fig:tem_itg}, the neural networks fits (solid and dashed lines) from the combined ITG + TEM networks accurately reproduces the non-trivial structure of the ITG-TEM transition. 

\section{Physics-based cost function}
\label{sec:cost_function}
Training a neural network means optimizing the weights and biases of the network to minimize a cost function $C$, which typically compares for each set of inputs, the neural network output to desired targets - in our case the QuaLiKiz input-output mapping. Typically the cost function consists of a measure of goodness-of-fit, and a regularizing term, as already shown in Eq.~\ref{eq:costfunction}. We have customized the cost function for our application beyond this standard implementation, to impose prior physics knowledge of the mapping structure into the system. This prior knowledge consists of: sharp instability thresholds, zero flux in the region where no instabilities are predicted, and identical transport flux thresholds for all transport channels. This last point has been treated through the leading-flux and flux-ratio paradigm introduced in section \ref{sec:training_targets}. We now summarize the other two.

The sharpness of the critical threshold is achieved by only including the unstable points (where instabilities are predicted) in the measure of goodness-of-fit $C_{good}$ for the leading flux regression. Otherwise, if including the zero flux points explicitly, then due to regularization some smoothing at the discontinuous critical threshold region is inevitable, leading to a loss of accuracy in the regression. By only including the unstable points, the leading flux neural network predictions are free to extrapolate to negative fluxes below the critical threshold, which are then clipped to zero in the transport model implementation, leading to the desired sharp critical threshold behaviour for all transport channels.

We then wish to avoid any possible FFNN extrapolation to spurious non-zero fluxes in the stable region below critical threshold. This is done by controlling the allowed range of extrapolation in the stable region. We add an additional penalty term $C_{stab}$ in the cost function for the leading flux regression, for samples predicted to be stable by QuaLiKiz. This penalty term punishes positive FFNN predictions in ostensibly stable regions, while remaining zero for negative FFNN predictions in the stable region (which are then subsequently clipped to zero).

The customized cost function is summarized in Eq.\ref{eq:customizedCF}.  The free parameters $\lambda_{regu}$, $\lambda_{stab}$, and $c_{stab}$, as well as other hyperparameters like network topology, are then optimized using a simple grid search. To test generalization, the dataset is split in a test set of 5\% never seen during training, and a validation set of 5\% used during training to avoid overfitting on training data. The remainder is used as training set. So, for each network prediction $NN_i$ relating to a QuaLiKiz calculation $QLK_i$ we have for all $n$ samples and $k$ weights:
\begin{equation}
\label{eq:customizedCF}
    C = C_{good} + \lambda_{regu} C_{regu} + \lambda_{stab} C_{stab}
\end{equation}
\newcommand{\mli}[1]{\mathit{#1}}
\begin{align}
    C_{good} &=
        \begin{cases}
        \frac{1}{n} \sum\limits^n_{i=1} (\mathit{QLK}_i - \mathit{NN}_i)^2,& \text{if } \mathit{QLK}_i \neq 0 \\
        0, & \text{if } \mathit{QLK}_i = 0
        \end{cases} \\
    C_{regu} &= \sum\limits^k_{i=1} w_i^2 \\
    C_{stab} &=
        \begin{cases}
        0, & \text{if } \mathit{QLK}_i \neq 0 \\
        \frac{1}{n} \sum\limits^n_{i=1} \mathit{NN}_i - c_{stab},& \text{if } \mathit{QLK}_i = 0
        \end{cases}
\end{align}
The final values of the free parameters of this hyperparameter optimization exercise can be found in Table \ref{tab:hyperpar}.

\section{Network training set data filtering}
\label{sec:filtering}
Inaccurate data in the training set can have a deleterious impact on the neural network training by overly biasing the regression towards an inaccurate representation. Such inaccuracies can arise due to unexplored corners in parameter space present in the QuaLiKiz scan, outside the commonly used (and experimentally relevant) parameter regimes of the code. While several code improvements were already made for some of these regimes on a case by case basis, surveyal by eye of the entire dataset was not feasible due to database size. In addition, due to the relatively low accuracy (2\%) demanded on the internal QuaLiKiz cubature routines to increase calculation speed, numerical errors related to occasional underestimation of integration relative accuracy can lead to spurious flux calculations. Therefore a conservative approach was taken in filtering the training set to remove untrusted QuaLiKiz flux calculations. As the dataset is generally too large for memory, the dask framework\cite{dask} was used to allow for general out-of-core processing of arbitrarily large array-like structures. For the networks trained in this work, data points were deemed untrusted and not included in training, according to the following heuristic criteria, which all indicate an internal integration routine might have failed. The percentage of the dataset filtered at each step is indicated in the list below. The dataset was generated with QuaLiKiz v2.4.0. Due to continuous improvements these numbers may be decreased in later versions. Each filter is applied consecutively, so multiple filters might filter out the same sample. The quoted percentages are with respect to the total dataset size.

\begin{itemize}
  \item Difference between total particle flux and derived particle flux from diffusion and convection transport coefficients is more than 50\%, i.e. $\left\lvert \frac{\Gamma_s - \left(-D_s dn_s/dr + V_s n_s\right)}{\Gamma_s}\right\rvert  > 0.5 $ (2.32\%)
  \item Total heat flux was negative (3.92\%)
  \item Difference between unweighted sum of ITG + TEM mode contributions, and self-consistent total flux calculation, was more than 50\% (0.08 \%)
  \item Ambipolarity was violated by more than 50\%. (1.75 \%). Note that the QuaLiKiz dispersion relation solution is intrinsically ambipolar, but cases of reduced convergence in the separate quasilinear flux integrations for ions and electrons can occasionally lead to a loss of ambipolarity. Transport solvers solve for either the electrons or the ions, assuming ambipolarity, so this occasional loss in transport model ambipolar ouputs for isolated calls doesn't lead to a loss of ambipolarity in practice. 
  \item Any transport coefficient is non-zero but predicted to be smaller than $10^{-4}$ in GyroBohm (GB) units (3.14 \%)
\end{itemize}
To increase prediction quality in experimentally relevant regimes, all points with either total ion or electron energy fluxes larger than 33 (in GB units with the minor radius as length scale) were removed (6.6\%), which is far beyond typically encountered in core plasmas. Additionally it was found essential for the flux-ratio predicting networks to remove low and high fraction values. These were removed by visual examination of the data histograms and determining cut-off points corresponding to tails of the distributions. With the percentage of points dropped in brackets, these were determined as:
  \begin{align}
    0.05 < q_{e,ITG} / q_{i,ITG} < 1.5 (0.73\%)\\
    0.02 < \lvert\Gamma_{e,ITG} / q_{i,ITG}\rvert < 0.6 (1.83\%)\\
    0.05 < q_{i,TEM} / q_{e,TEM} < 2.0 (24.06\%)\\
    0.03 < \lvert\Gamma_{e,TEM} / q_{e,TEM}\rvert < 0.8 (69.52\%)
  \end{align}
Note that the TEM filtering bounds are likely too strict, and will be improved upon in further work. None of the plasmas shown in Section \ref{sec:integrated_modelling} to test QLKNN within integrated modeling display TEM modes.

\section{Physics-based measures of goodness}
\label{sec:measures}
\begin{figure}
  \includegraphics[width=\linewidth]{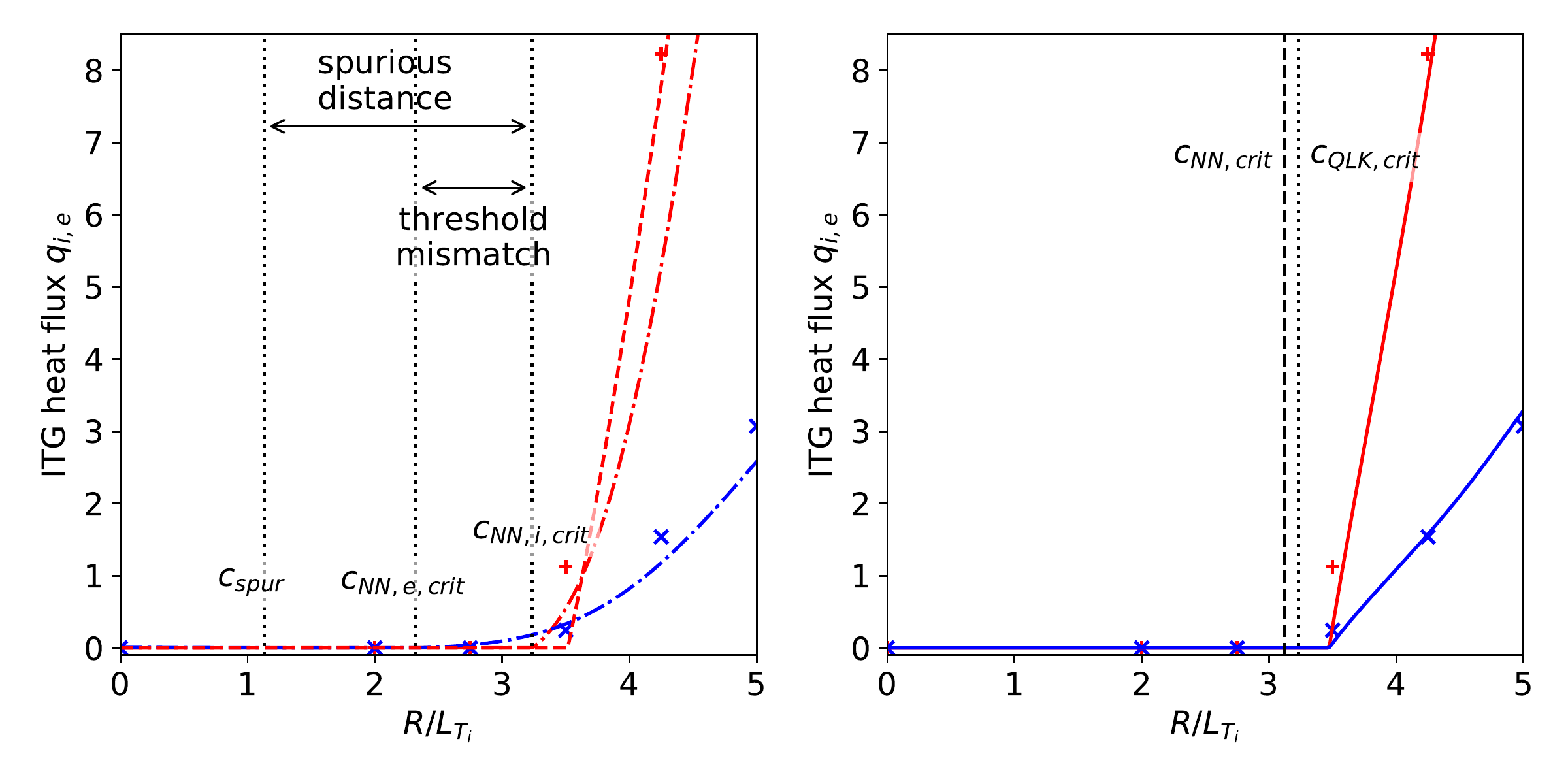}
  \caption{\label{fig:non_optimized_network}Predictions of the ITG driven heat flux for the ions (red) and electrons (blue). We show three types of networks, networks trained using a standard RMS error on both the stable and unstable points (dash-dot, left) and only on the unstable points (dashed, left). These networks show clear mismatch between transport channels and QuaLiKiz prediction, as well as a small but finite prediction of fluxes in the stable region. The physics-based neural network (right) have no mismatch between transport channels, a sharper threshold closer to the QuaLiKiz prediction, as well as no prediction of flux in the stable region.}

\end{figure}

Performance indicators are a critical tool for differentiating the quality of different neural networks, trained with different hyperparameters, to assess the optimal networks to use in our application. In contrast to classical regression tasks, the \textit{final loss} - meaning the final value of the optimized cost function - is not the key performance indicator of the trained network. Instead, for our application, how well the trained neural network performs as a transport model within integrated transport modelling is the most important. However, using the integrated model directly in the training pipeline is cumbersome and computationally expensive. Instead, we define metrics relating to the aforementioned capture of known physical features and use these in conjunction with the classical test loss to judge the quality of the trained networks after training. To do this we take 1D slices in the main driving gradient (for each mode) from the full dataset and let the network predict over the range of this slice. The main driving gradients are taken to be the electron temperature gradient for TEM and ETG, and the ion temperature gradient for ITG. The full dataset contains \num{2.4e7} such slices, but taking 5\% was sufficient to statistically differentiate networks with different hyperparameters. We first define for each slice:

\begin{itemize}
    \item The \textit{neural network critical gradient} $c_{NN,crit}$; the location where the neural network leading flux predictions cross from positive to negative fluxes corresponding to the transition from unstable to stable QuaLiKiz regions.  
    \item The \textit{spurious stable prediction} $c_{spur}$; the first encountered point in the QuaLiKiz stable region, when descending from high gradient to low gradient, where the neural network predictions spuriously transitions from negative flux (clipped to zero in the transport code implementation) to positive flux
    \item The \textit{QuaLiKiz critical gradient proxy} $c_{QLK,crit}$; the midpoint between the gradient slice gridpoints corresponding to the transition from zero to positive fluxes in the original QuaLiKiz data
\end{itemize}
Using these quantities, we found the following measures to be important:
\begin{itemize}
  \item The \textit{no threshold fraction}; the percentage of slices where QuaLiKiz predicts a threshold (e.g. has a zero-flux-crossing), and QLKNN does not
    \item The \textit{spurious flux fraction}; the percentage of slices where QLKNN predicts spurious flux in the stable region
    \item The \textit{threshold misprediction}; the mean absolute distance between the QuaLiKiz and QLKNN thresholds $\frac{1}{n}\sum_i^n \lvert c_{NN,crit} - c_{QLK,crit} \rvert $
    \item The \textit{threshold mismatch}; the mean absolute distance between the predicted thresholds of two transport channels, for example between ions and electrons $\frac{1}{n}\sum_i^n \lvert c_{NN,e,crit} - c_{NN,i,crit} \rvert$. Necessarily zero for the QLKNN methodology.
    \item The \textit{unstable zone smoothness}; the smoothness in the unstable zone as defined from the second derivative with respect to the driving gradient: $\frac{1}{n}\sum_i^n\lvert \frac{\partial^2 x}{\partial \left(R/LT_s\right)^2 }\rvert, \text{if $R/LT_s > c_{NN,crit}$}$. This strongly depends on the regularisation hyperparameter.
    \item The \textit{spurious distance}, or the relative distance of spurious stable flux prediction and predicted threshold $\frac{c_{NN,crit} - c_{spur}}{c_{NN,crit}}$
\end{itemize}
\begin{table}
  \caption{\label{tab:measures} An overview of the measures of goodness as described in Section \ref{sec:measures}: The percentage of slices with no threshold (No thresh frac) and without spurious flux predictions (No spurious frac), the absolute threshold mismatch (Abs thresh mismatch), relative spurious flux prediction distance (Rel spurious dist), and smoothness in the unstable zone (Unstab zone smooth). These quantities are shown for the three leading flux networks $q_{e,ETG}$, $q_{e,TEM}$, and $q_{i,ITG}$. These statistics were taken on a reduced 7D dataset, fixing $Z_{eff}=1$ and $\nu^*=1e-3$. No attempt is made to combine these measures of goodness into a single final value, nor is currently known what the upper and lower bounds are. However, as shown later in this work, these values were found sufficient for good model performance in integrated modelling. The RMS error was calculated on the unstable 9D test set with values higher than 33 filtered out.}
  \begin{ruledtabular}
    \begin{tabular}{p{0.11\linewidth} *{6}{p{0.134\linewidth}}}
    {} & \raggedright RMS test [GB] & \raggedright No thresh frac [\%] & \raggedright No spurious frac [\%] & \raggedright Abs thresh mismatch & \raggedright Rel spurious dist [\%] & Unstab zone smooth \\
      \hline
    $q_{e,ETG}$ & 2.0 & 3.3 & 97.7 & -0.38 & -0.44 & 0.017 \\
    $q_{e,TEM}$ & 1.8 & 14.3 & 98.6 & -0.31 & -0.70 & 0.008 \\
    $q_{i,ITG}$ & 2.3 & 4.2 & 99.2 & -0.26 & -0.52 & 0.0300
\end{tabular}
  \end{ruledtabular}
\end{table}
An overview of these distances is shown in Figure \ref{fig:non_optimized_network}. A trained network never has an absolute minimum in all these metrics simultaneously, so instead a trade-off is made. In this work we have not attempted to unify these metrics in a single value. Instead, the metrics are used as guidance to select a small number of networks that are then tested inside the integrated model. This adds some potential bias to the process, and future work would profit from investigation of objective and quantitative measures of goodness for trained networks. The metrics for the final implemented networks can be found in Table \ref{tab:measures}. All these metrics for measures of goodness end up very similar both for the leading flux network and their associated ratio network because of the choice of training targets described in Section \ref{sec:training_targets}. Finding minimum required values of the metrics is outside the scope of this work, but we note the low percentage of stable flux predictions for all networks, and the low values for threshold mismatch. Because of the low percentage of stable flux predictions, the relative spurious distance is thought to be of less importance, while the smoothness was assumed to be sufficient by visual inspection of many neural network predictions on random slices. Finally, while the measures of goodness for the TEM networks are not as good as the others, we found it encouraging enough to implement them in the later-described transport models. However, for future work improving these networks specifically would be beneficial.

\section{Neural network training pipeline}
\label{sec:pipeline}
The networks were trained using the TensorFlow\cite{tensorflow} framework. TensorFlow is an open source framework allowing various machine learning algorithms to be run efficiently on heterogeneous machines, both for CPU and for GPU architectures. The framework can be used to train neural networks out-of-the-box, but as it is a general framework care has to be taken that the use-case it is applied to matches the expectations of the framework. In this work, we have identified and worked around two limitations of the TensorFlow framework at time of writing. Firstly, TensorFlow is most commonly used for deep learning. In deep learning, the amount of training samples versus the size of the network, and thus its evaluation speed, is relatively small. In this work, the networks are shallow and the amount of training samples large. As such, we have implemented a simple but shuffling algorithm using numpy\cite{numpy}, which is a factor 2 faster for this application. This results in 1.25x (CPU) to 2x (Tesla P100 GPU) reduction of training time. Secondly, TensorFlow uses its own proprietary format to save the trained neural network weights and biases to disk. This would mean any integrated framework would need to depend on TensorFlow/python to use the neural network predictions. This is inconvenient and non-performant for many integrated frameworks, especially if they are in MATLAB (RAPTOR) and Fortran (JINTRAC). Instead, we wrote a lightweight communication format using JSON between TensorFlow, and a re-implementation of the network in Fortran with wrappers for Python and MATLAB. Using these MKL accelerated native Fortran functions, the baseline QLKNN model, 7 networks with 3 layers and 128 neurons each for 24 radial points, can be evaluated within \SI{1.4}{m\second} on a single core or \SI{60}{m\second} if the derivatives of the neural network output with respect to the neural network inputs are also evaluated. This can be accelerated to \SI{0.3}{ms} and \SI{9}{ms} respectively by parallelizing over 7 cores using MPI. These timings were obtained on a Intel(R) Xeon(R) CPU E5-2665 0 @ 2.40GHz. The FFNN and QLKNN wrappers are freely available at GitLab.com\cite{qlknn_fortran_repository}.

Neural network training involves optimizing training hyperparameters. While many algorithms to optimize hyperparameters exist, the authors are not aware of a commonly used readily-available framework to do this. As such, we have written a thin wrapper around TensorFlow, using a PostgreSQL database and Spotify's Luigi framework\cite{luigi_repository} to interact with supercomputer job schedulers and train, validate and analyse networks trained with the QLKNN training framework. This allows to set up simple hypergrid hyperparameter scans. For the dataset used in this work we have found optimal hyperparameters which work well for a dataset of reduced 7D space (fixing $Z_{eff}=1$ and $\nu^*=1e-3$), also work sufficiently well for networks trained on the full 9D space. Additionally, hyperparameters optimal for ITG neural networks were found to work well for networks for the other modes in exploratory hyperparameter scans. This brings down the training time from 24 hours per network to approximately an hour. Using this property, we have done wide scans of the following hyperparameters for $q_{i,ITG}$, resulting in over 1000 trained neural networks:

\begin{table}
  \caption{\label{tab:hyperpar}Results of hyperparameter optimization. All networks were trained with the ADAM algorithm with learning rate $\alpha = 0.001$ and decays $\beta_1 = 0.9$ and $\beta_2 = 0.999$, not optimized in this work. A test set of 5\% was kept separate during this optimization.}
  \begin{ruledtabular}
    \begin{tabular}{*{1}{l} *{1}{r}}
      Variable & Optimized value \\
      \hline
      Number of layers & 3 \\
      Neurons per layer & 128 \\
      Early stopping patience & 15 \\
      L2 regularization strength $\lambda_{regu}$ & 1e-5 \\
      Stability positive scale $\lambda_{stab}$ & 1e-3 \\
      Stability positive barrier $C_{stab}$ & 5 \\
      Validation fraction & 5\% \\
    \end{tabular}
  \end{ruledtabular}
\end{table}
This resulted in the optimal hyperparameters shown in Table \ref{tab:hyperpar}. All networks were trained with RMSE as measure of goodness and L2 and early stopping as regularization, using the ADAM algorithm\cite{kingma2014adamoptimizer}. The hyperparameters were found by doing a wide scan for the leading fluxes on a reduced 7D dataset. These were then refined by a smaller scan on the full 9D dataset, resulting in the same optimal 7D values. The obtained network weights and biases are freely available on GitLab\cite{qlknn_hyper_repository}.
\section{Simulation settings}
In this chapter we show relevant JINTRAC and RAPTOR settings for all runs in Table \ref{tab:jetto_settings} and the associated run timing in Table \ref{tab:jetto_timing}. The isotope mix of the non-benchmark simulations can be found in Table \ref{tab:jetto_settings_imp}. Finally, the used proxy-sawtooth transport patches can be found in Tables \ref{tab:jetto_settings1}, \ref{tab:jetto_settings2}, and \ref{tab:jetto_settings3}.
\label{sec:full_jetto_settings}
\begin{table*}
  \caption{\label{tab:jetto_settings} JINTRAC and RAPTOR settings of simulations in this paper. The same settings were used for the QLKNN and QuaLiKiz simulations. All simulations have lower limits of $10^{-4} m^2 s^{-1}$ for thermal diffusion, negligible in the turbulent transport region.}
  \begin{ruledtabular}
    \begin{tabular}{*{1}{l} *{5}{r}}
      Field name/option & Value/setting & & & \\
      Shot number & 73342 & 92398 & 92436 & 92436 (bench) & 92436 (RAPTOR)\\
      Number of grid points & 51 & 101 & 101 & 25 & 9 \\
      Start time a (s) & 20.75 & 6.3779 & 10 & 10 & 10 \\
      End time a (s) & 22.75 & 8.6 & 12 & 12 & 12 \\
      Min. time step (s) & 1e-13 & 1e-8 & 1e-13 & 1e-13 & 0.1 \\
      Max. time step (s) & 1e-3 & 1e-3 & 1e-3 & 1e-3 & 0.1\\
      Main ion mass (u) & 2 & 2 & 2 & 2 & 2 \\
      Simulation boundary ($\rho_{N,tor}$)& 0.85 & 0.85 & 0.85 & 0.85 & 0.85\\
      Equilibrium & EFIT & ESCO & ESCO & EQDSK & EQDSK\\
      Equilibrium boundary & - & 0.995 & 0.998 & - & - \\
      Toroidal field (for ESCO) & - & 2.798 & 2.8 & - & - \\
      Plasma current (A) & 2.5e6 & 2.2e6 & 2.9e6 & 2.9e6 & 2.9e6 \\
      Neoclassical transport model & NCLASS & NCLASS & NCLASS & - & - \\
      Bootstrap current & yes & yes & yes & yes & yes \\
      Particle transport min ($cm^2/s$) & 10 & 1 & 10 & 10 & 10 \\
      Impurities & Interpretive & Interpretive & SANCO & - & - \\
      Numerical scheme & Predictor-Corrector & Predictor-Corrector & Predictor-Corrector & Predictor-Corrector & Implicit \\
    \end{tabular}
  \end{ruledtabular}
\end{table*}
%
%
\begin{table*}
  \caption{\label{tab:jetto_timing}Exact run times for the simulations in this paper. Also shows the amount of transport model (QLKNN or QLK) evaluations for all runs. This is always 2* the amount of timesteps in JINTRAC, but depends on the number of newton evaluations needed in RAPTOR. Usually, as soon as a physically consistent state is reached, this is 2 or 3 per time step.
  }
  \begin{ruledtabular}
    \begin{tabular}{*{2}{l} *{10}{r}}
      &Shot number & \multicolumn{2}{c}{73342} & \multicolumn{2}{c}{92398} & \multicolumn{2}{c}{92436} & \multicolumn{2}{c}{92436 (bench)} & \multicolumn{2}{c}{92436 (RAPTOR)}\\
                   &                       & QLK  & QLKNN & QLK  & QLKNN & QLK   & QLKNN & QLK   & QLKNN & \multicolumn{2}{r}{QLKNN} \\
                   & Amount of cores       & 16   & 2     & 16   & 2     & 16    & 2     & 16    & 1     & \multicolumn{2}{r}{1} \\
      Rotationless & number of evaluations & 5788 & 4002  & 4454 & 4456  & 19934 & 4076  & 4006  & 4006  & \multicolumn{2}{r}{51}\\
      Rotation     & number of evaluations & 4654 & 4002  & 4452 & 4506  & 4480  & 4040  & -     & -     & \multicolumn{2}{r}{-} \\
      Rotationless & runtime (m)           & 284  & 1     & 775  & 8     & 8550  & 34    & 217   & 0.2   & \multicolumn{2}{r}{0.1}  \\
      Rotation     & runtime (m)           & 1564 & 1     & 644  & 8     & 11708 & 33    & -     & -     & \multicolumn{2}{r}{-}
    \end{tabular}
  \end{ruledtabular}
\end{table*}
\begin{table}
  \caption{\label{tab:jetto_settings_imp} Used impurities in the non-benchmark JINTRAC-QuaLiKiz and JINTRAC-QLKNN runs. The table shows the mass (m), charge (c) and amount of super stages (ss) used in SANCO if applicable.}
  \begin{ruledtabular}
    \begin{tabular}{*{1}{l} *{9}{r}}
      & \multicolumn{3}{c}{species 1} & \multicolumn{3}{c}{species 2} & \multicolumn{3}{c}{species 3} \\
      simulation & m & c & ss & m & c & ss & m & c & ss \\
      73342 & 12 & 6 & - \\
      92398 & 58 & 28 & - \\
      92436 & 9 & 4 & 4 & 58 & 28 & 28 & 184 & 74 & 74 \\
    \end{tabular}
  \end{ruledtabular}
\end{table}
%
\begin{table*}
  \caption{\label{tab:jetto_bgb} Extra BgB transport in JETTO. Not implemented in RAPTOR, and negligible in the predicted turbulent unstable region.}
  \begin{ruledtabular}
    \begin{tabular}{*{1}{l} *{4}{r}}
      Field name/option & Value/setting & & & \\
      Shot number & 73342 & 92398 & 92436 & 92436 (bench) \\
      Particle diffusion multiplier & 1 & 1 & 1 & 0 \\
      Thermal diffusion multiplier (bohm, electron) & 0.08 & 0.03 & 0.08 & 0 \\
      Thermal diffusion multiplier (bohm, ion) & 0.08 & 0.03 & 0.08 & 0 \\
      Momentum diffusion (prandl number) & 1 & 3 & 1.25 & 0 
    \end{tabular}
  \end{ruledtabular}
\end{table*}
\newcommand{\tcenter}{Centre ($\rho_{N,tor}$)}
\newcommand{\theight}{Height ($m^2 s^{-1}$)}
\newcommand{\twidth}{$2\sigma$ width ($\rho_{N,tor}$)}
\begin{table}
  \caption{\label{tab:jetto_settings1} Sawtooth-proxy core transport patch for JET \#73342}
  \begin{ruledtabular}
    \begin{tabular}{*{1}{l} *{3}{r}}
      &\multicolumn{3}{c}{Additional transport} \\
       & $\chi_i$ & $\chi_e$ & $D_i$ \\
      Shape & Gaussian & Gaussian & Gaussian \\
      \tcenter & 0 & 0 & 0 \\
      \theight & 4 & 2 & 2 \\
      \twidth & 0.25 & 0.25 & 0.25
    \end{tabular}
  \end{ruledtabular}
\end{table}
\begin{table}
  \caption{\label{tab:jetto_settings2} Additional core transport patch for JET \#92398}
  \begin{ruledtabular}
    \begin{tabular}{*{1}{l} *{3}{r}}
      &\multicolumn{3}{c}{Additional transport} \\
       & $\chi_i$ & $\chi_e$ & $D_i$ \\
      Shape & - & Gaussian & - \\
      \tcenter & - & 0 & - \\
      \theight & - & 0.1 & - \\
      \twidth & - & 0.15 & - \\
    \end{tabular}
  \end{ruledtabular}
\end{table}
\begin{table}
  \caption{\label{tab:jetto_settings3} Sawtooth-proxy core transport patch for JET \#92436}
  \begin{ruledtabular}
    \begin{tabular}{*{1}{l} *{3}{r}}
      &\multicolumn{3}{c}{Additional transport} \\
       & $\chi_i$ & $\chi_e$ & $D_i$ \\
      Shape & Gaussian & Gaussian & Gaussian \\
      \tcenter & 0 & 0 & 0 \\
      \theight & 2 & 1 & 1 \\
      \twidth & 0.212 & 0.212 & 0.212 \\
    \end{tabular}
  \end{ruledtabular}
\end{table}
\begin{table*}
  \caption{\label{tab:code_version} Equivalent code version used in this work, as determined by \texttt{git describe -{}-abbrev=6 -{}-tags}. All results in this paper should be reproducible by using the code versions in this table.}
  \begin{ruledtabular}
    \begin{tabular}{*{1}{l} *{2}{r}}
    part & version & repository \\
    QLKNN-fortran & v1.0.0 & \url{https://gitlab.com/qualikiz-group/QLKNN-fortran} \\
    QLKNN networks & v0.5.0-1-ge2c20a & \url{https://gitlab.com/qualikiz-group/qlknn-hyper} \\
    QLKNN tools and filters & v1.0.0 & \url{https://gitlab.com/Karel-van-de-Plassche/QLKNN-develop} \\
    Physics-uninformed networks & 0.1.0 & \url{https://gitlab.com/qualikiz-group/qlknn-fullflux} \\
    QuaLiKiz & v2.6.1-5-g95d8df & \url{http://qualikiz.com} \\
    QuaLiKiz (dataset generation) & v2.4.0 & \url{http://qualikiz.com} \\
    JETTO & Release-v191219 & \url{https://git.ccfe.ac.uk/jintrac/jetto-sanco} \\
    RAPTOR & plassche\_PoP2019 & \url{https://gitlab.epfl.ch/spc/raptor}
    \end{tabular}
  \end{ruledtabular}
\end{table*}
\begin{table*}
  \caption{\label{tab:runs} Archived JETTO runs displayed in this paper.}
  \begin{ruledtabular}
    \begin{tabular}{*{1}{l} *{5}{r}}
    figure & line & user & shot & archive date & catalogue seq \\
    \ref{fig:jintrac_raptor_92436}  & JINTRAC-QLKNN                   & kplass  & 92436 & dec2919 & seq\#4 \\
                                    & JINTRAC-QuaLiKiz                & kplass  & 92436 & dec3019 & seq\#1 \\
   \ref{fig:nn_comp_92436}          & JINTRAC-QLKNN (rotationless)    & kplass  & 92436 & dec2919 & seq\#1 \\
                                    & JINTRAC-QLKNN (full-physics)    & kplass  & 92436 & dec2919 & seq\#2 \\
                                    & JINTRAC-QuaLiKiz (rotationless) & jcitrin & 92436 & may2419 & seq\#1 \\
                                    & JINTRAC-QuaLiKiz (full-physics) & jcitrin & 92436 & may2419 & seq\#2 \\
    \ref{fig:nn_comp_73342}         & JINTRAC-QLKNN (rotationless)    & kplass  & 73342 & dec2919 & seq\#1 \\
                                    & JINTRAC-QLKNN (full-physics)    & kplass  & 73342 & dec2919 & seq\#2 \\
                                    & JINTRAC-QuaLiKiz (rotationless) & jcitrin & 73342 & apr2319 & seq\#1 \\
                                    & JINTRAC-QuaLiKiz (full-physics) & jcitrin & 73342 & apr2319 & seq\#2 \\
   \ref{fig:nn_comp_92398}          & JINTRAC-QLKNN (rotationless)    & kplass  & 92398 & dec2919 & seq\#1 \\
                                    & JINTRAC-QLKNN (full-physics)    & kplass  & 92398 & dec2919 & seq\#2 \\
                                    & JINTRAC-QuaLiKiz (rotationless) & jcitrin & 92398 & jan2919 & seq\#1 \\
                                    & JINTRAC-QuaLiKiz (full-physics) & jcitrin & 92398 & oct1618 & seq\#1
    \end{tabular}
  \end{ruledtabular}
\end{table*}
\end{document}